%% file: main.tex
\documentclass[10pt,journal,compsoc]{IEEEtran}
\ifCLASSOPTIONcompsoc
  \usepackage[compress]{cite}
\else
  \usepackage{cite}
\fi
\ifCLASSINFOpdf
\else
\fi
\usepackage{amsmath}
\usepackage{graphicx,psfrag,epsf}
\usepackage{enumerate}
\usepackage{url} % not crucial - just used below for the URL 
\usepackage{hyperref}
\usepackage{url}
\usepackage[bottom]{footmisc}
\usepackage{wrapfig}

\usepackage{booktabs}  
\usepackage{graphicx}
\usepackage{amsmath}
\usepackage[caption=false]{subfig}
\usepackage{bm}
\usepackage{xcolor}
\usepackage[normalem]{ulem}
\useunder{\uline}{\ul}{}
\usepackage{amsthm}
\newtheorem{theorem}{Theorem}[section]
\newtheorem{lemma}[theorem]{Lemma}

\DeclareMathOperator*{\argmin}{arg\,min}
\usepackage{amsfonts}
\newcommand{\linebreakand}{%
  \end{@IEEEauthorhalign}
  \hfill\mbox{}\par
  \mbox{}\hfill\begin{@IEEEauthorhalign}
}

\begin{document}
\title{A Hierarchical Block Distance Model for Ultra Low-Dimensional Graph Representations}
\author{Nikolaos Nakis, Abdulkadir 
\c{C}elikkanat, Sune Lehmann, Morten Mørup}

\IEEEtitleabstractindextext{%
\begin{abstract}
Graph Representation Learning (GRL) has become central for characterizing structures of complex networks and performing tasks such as link prediction, node classification, network reconstruction, and community detection. Whereas numerous generative GRL models have been proposed, many approaches have prohibitive computational requirements hampering large-scale network analysis, fewer are able to explicitly account for structure emerging at multiple scales, and only a few explicitly respect important network properties such as homophily and transitivity. This paper proposes a novel scalable graph representation learning method named the Hierarchical Block Distance Model (HBDM). The HBDM imposes a multiscale block structure akin to stochastic block modeling (SBM) and accounts for homophily and transitivity by accurately approximating the latent distance model (LDM) throughout the inferred hierarchy. The HBDM naturally accommodates unipartite, directed, and bipartite networks whereas the hierarchy is designed to ensure linearithmic time and space complexity enabling the analysis of very large-scale networks. We evaluate the performance of the HBDM on massive networks consisting of millions of nodes. Importantly, we find that the proposed HBDM framework significantly outperforms recent scalable approaches in all considered downstream tasks. Surprisingly, we observe superior performance even imposing ultra-low two-dimensional embeddings facilitating accurate direct and hierarchical-aware network visualization and interpretation.
\end{abstract}

}

\maketitle

\IEEEdisplaynontitleabstractindextext

% \fi
\IEEEpeerreviewmaketitle

\input{1-introduction}
\input{2-methods}

\input{4-experiments}

\input{5-discussion}
%\clearpage
%\input{6-conclusion}
%\input{newmotivation}
% \input{notes}
%\input{6-appendix}

%\clearpage


\ifCLASSOPTIONcompsoc
  \section*{Acknowledgments}
\else
  \section*{Acknowledgment}
\fi

We would like to express sincere appreciation and thank the reviewers for their constructive feedback and their insightful comments. We would also like to thank Louis Boucherie, Lasse Mohr Mikkelsen, and Giorgio Giannone for the valuable and fruitful discussions. The authors gratefully acknowledge the Independent Research Fund Denmark for supporting this work [grant number: 0136-00315B].

% or


\ifCLASSOPTIONcaptionsoff
  \newpage
\fi



%\bibliographystyle{IEEEtran}

% \bibliographystyle{plain}
% \bibliographystyle{agsm}

\bibliographystyle{IEEEtran}
\bibliography{Bibliography-MM-MC}

\begin{IEEEbiography}[{\includegraphics[width=1in,height=1.25in,clip,keepaspectratio]{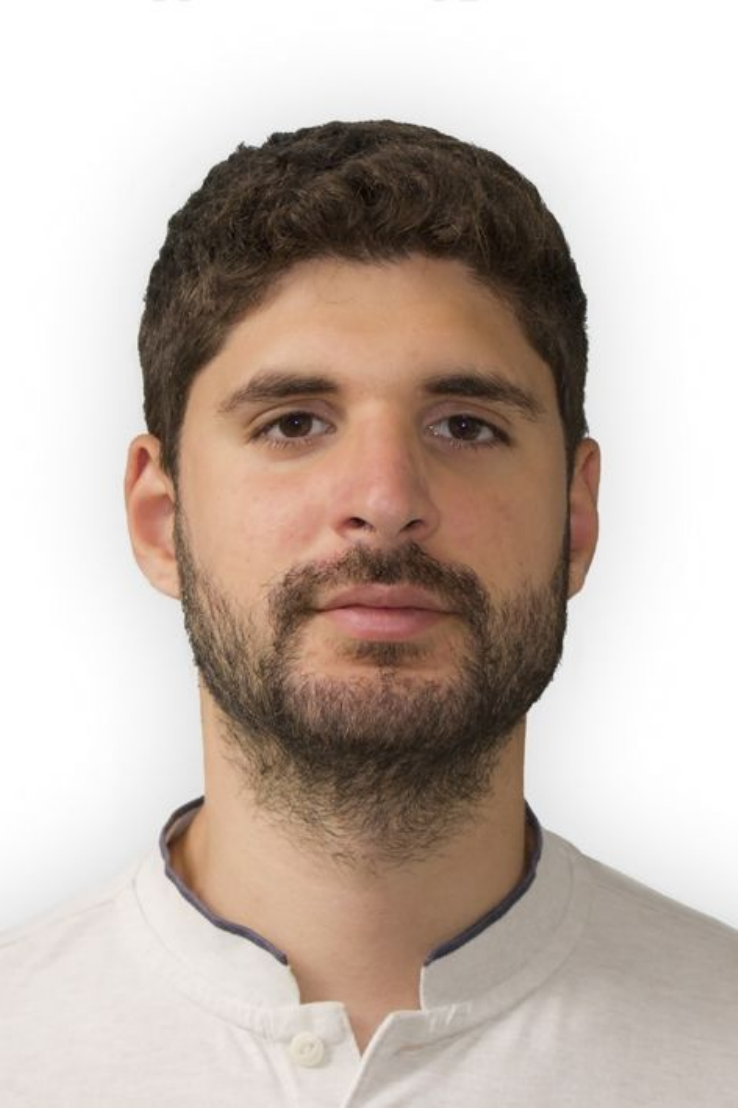}}]{Nikolaos Nakis}
is currently a Ph.D. student at the Section for Cognitive Systems of the Technical University of Denmark. He received his BS in physics from the National and Kapodistrian University of Athens and his MS degree in mathematical modeling and computation from the Technical University of Denmark. His research mainly focuses on machine learning applied to complex systems and graph representation learning.
\end{IEEEbiography}

\begin{IEEEbiography}[{\includegraphics[width=1in,height=1.25in,clip,keepaspectratio]{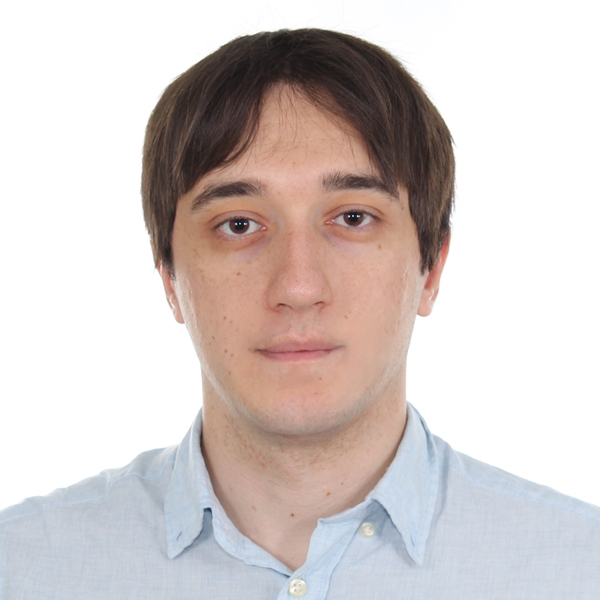}}]{Abdulkadir \c{C}elikkanat} is currently a postdoctoral researcher at the Section for Cognitive Systems of the Technical University of Denmark. He completed his Ph.D. at the Centre for Visual Computing of CentraleSupélec, Paris-Saclay University, and he was also a member of the OPIS team at Inria Saclay. Before his Ph.D. studies, he received his Bachelor degree in Mathematics and Master's degree in Computer Engineering from Bogazi\c{c}i University. His research mainly focuses on the analysis of graph-structured data. In particular, he is interested in graph representation learning and its applications for social and biological networks.
\end{IEEEbiography}

\begin{IEEEbiography}[{\includegraphics[width=1in,height=1.25in,clip,keepaspectratio]{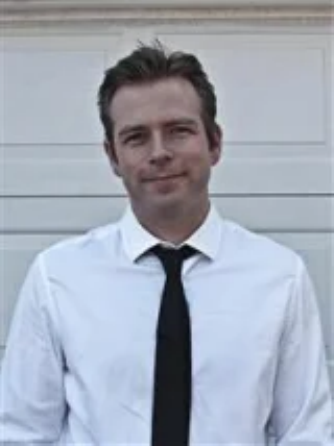}}]{Sune Lehmann}
Sune's work focuses on a quantitative understanding of social systems based on massive data sets. A physicist by training, his research draws on approaches from the physics of complex systems, machine learning, and statistical analysis. He works on large-scale behavioral data and while Sune's primary focus is on modeling complex networks, his research has made substantial contributions on topics such as human mobility, sleep, academic performance, complex contagion, epidemic spreading, and behavior on Twitter. He is the author of multiple high-impact papers and his research has won various prizes.
\end{IEEEbiography}

\begin{IEEEbiography}[{\includegraphics[width=1in,height=1.25in,clip,keepaspectratio]{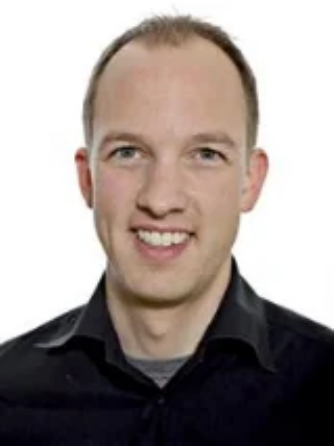}}]{Morten Mørup}
received the MS and PhD degrees
in applied mathematics from the Technical University of Denmark, where he is currently
professor at the Section for Cognitive Systems at
DTU Compute. He has been an associate editor of the
IEEE Transactions on Signal Processing and his
research interests include machine learning, neuroimaging, and complex network modeling.
\end{IEEEbiography}

\end{document}

%% file: 1-introduction.tex
\IEEEraisesectionheading{\section{Introduction}}
\IEEEPARstart{N}etworks naturally arise in a plethora of scientific areas to model interactions between entities from physics to sociology and biology, with many instances such as collaboration, protein-protein interaction, and brain connectivity networks \cite{newman} to mention but a few. In recent years, Graph Representation Learning (GRL) approaches have attracted great interest due to their outstanding performance compared to classical techniques for arduous problems such as link prediction \cite{libennowel-cikm03}, node classification \cite{srl2007,node2vec-kdd16}, and community detection \cite{FORTUNATO201075}.

%Many existing GRL methods \cite{GRL-survey-ieeebigdata20} mainly aim to capture the underlying intrinsic relationships among the nodes by either performing random walks \cite{deepwalk-perozzi14, node2vec-kdd16} over the network or designing a matrix capturing the first or higher order node proximities \cite{cao2015grarep, HOPE-kdd16}. However, they require high computational and space costs reluing either on exact node sampling procedures or expensive factorizations of dense proximity matrices. The recent Graph Neural Networks (GNNs) \cite{GRL-survey-ieeebigdata20, GRL-survey-ieeebigdata20} provide effective tools in learning the node representations by leveraging the side information such as node attributes; nevertheless, they also face computational difficulties, especially for large-scale networks consisting of millions of nodes and edges. Although recent studies aim to alleviate the computational burden of the algorithms through matrix sparsification tools \cite{netsmf-www2019} or hierarchical representations \cite{louvainNE-wsdm20, harp-aaai18}, the performance of the methods in downstream tasks significantly drops, and they require larger embedding sizes to compensate for the loss.

Numerous GRL methods have been proposed, see also   \cite{GRL-survey-ieeebigdata20} for a survey. The leading initial works are the random walk-based methods \cite{deepwalk-perozzi14, node2vec-kdd16, line, expon_fam_emb, biasedwalk}, leveraging the Skip-Gram algorithm \cite{MSCC+13} to learn the node representations. Matrix factorization-based algorithms \cite{cao2015grarep, GRL-survey-ieeebigdata20} have also become prominent, extracting the embedding vectors by decomposing a designed feature matrix. Furthermore, neural network models \cite{GRL-survey-ieeebigdata20, graphsage_hamilton} have been proposed for graph-structured data, returning outstanding performance by combining node attributes and network structure when learning embeddings. Recent studies \cite{prone-ijai19} aim to alleviate the computational burden of these algorithms through matrix sparsification tools \cite{netsmf-www2019}, hierarchical representations \cite{louvainNE-wsdm20, harp-aaai18}, or by fast hashing schemes \cite{randne-icdm18}. %the performance of the methods in downstream tasks significantly drops, and they require larger embedding sizes to compensate for this loss.

Latent Space Models (\textsc{LSM}s) for the representation of graphs have been quite popular over the past years \cite{past1,past2,past3,past4,past5,expl2,LSM_geo}, especially for social networks analysis \cite{recent_lsm1,recent_LSM2} facilitating community extraction \cite{soc2} and characterization of network polarization \cite{soc1}. \textsc{LSM}s utilize the generalized linear model framework to obtain informative latent node embeddings while preserving network characteristics. The choice of latent effects in modeling the link probabilities between the nodes leads to different expressive capabilities characterizing network structure. In particular, in the Latent Distance Model (\textsc{LDM}) \cite{exp1} nodes are placed closer in the latent space if they are similar or vice-versa. \textsc{LDM} obeys the triangle inequality and thus naturally represents transitivity \cite{hom1,hom0} ({\it"a friend of a friend is a friend"}) and network homophily \cite{hom2,hom3} ({\it a tendency where similar nodes are more likely to connect to each other than dissimilar ones}). Homophily is a very well-known and well-studied effect appearing in social networks \cite{hom1,hom2,hom3} and essentially describes the tendency for people to form connections with those that share similarities with themselves. Similarities can be drawn from meta-data (observed node attributes) and may refer to shared demographic properties, political opinions, etc.  Homophily has been observed among a broad range of collaborations (see \cite{hom0} for a complete overview). Homophily can also be accounted for based on the unobserved attributes as defined by the \textsc{LDM} as shown in \cite{KRIVITSKY2009204}. Homophily explains prominent patterns as expressed in social networks in terms of transitivity, as well as, balance theory (“the enemy of my friend is an enemy”) \cite{balance_theory}. More specifically, in an \textsc{LDM} we can extend the meaning of similarity to some unobserved (latent) covariates, i.e., latent embeddings $\mathbf{Z}$. The higher similarity between nodes translates here to a stronger relationship between the nodes and thereby a higher probability of observing connections. As a result, for two similar nodes $\{i,j\}$ the pairwise distance $|\mathbf{z}_i - \mathbf{z}_j|_2$ should be small which further implies that for a different node $\{k\}$ we obtain $|\mathbf{z}_i -\mathbf{z}_k|_2 \approx |\mathbf{z}_j - \mathbf{z}_k|_2$. The latter concludes that nodes $\{i,j\}$ are similar since they share similar relationships with the rest of the nodes. 

The approach has been extended to bipartite networks in \cite{Friel6629} by introducing mode-specific embedding vectors and community detection by endowing the \textsc{LDM} with a Gaussian Mixture Model prior to promoting cluster structures in the latent space forming the latent position clustering model (\textsc{LPCM})\cite{LPC,KRIVITSKY2009204}. %In the work of \cite{Friel6629}, the \textsc{LDM} was extended to the study and modeling of bipartite structures% where a network of Irish companies and its corresponding directors was analyzed
However, the \textsc{LDM} is unable to account for possible stochastic equivalence as defined by the Stochastic Blockmodels \cite{holland1983stochastic,doi:10.1198/016214501753208735}, i.e. ({\it"groups of nodes defined by shared intra- and inter-group relationships"}) defining non positive-semi-definite latent representations. The LSMs were advanced to characterize such stochastic equivalence by imposing an Eigenmodel admitting negative eigenvalues \cite{hoff2007modeling}. These latent space methods are attractive due to their simplicity, as they define well-structured inference problems and are characterized by high explanatory power \cite{LSM_geo}. The time and space complexities are their main drawbacks as the likelihood function scales by the number of node pairs (i.e., quadratically in the number of nodes for a unipartite graph) typically addressed using subsampling strategies \cite{Raftery}. 

Many real-world networks are composed of structures emerging at multiple scales which can be expressed using hierarchical representations \cite{hierarchicalrepr}. Several methods have thus been advanced to such hierarchical representations including stochastic block model approaches \cite{clauset2008hierarchical, roy2007learning, NIPS2008_fe8c15fe, herlau2012detecting, herlau2013modeling,Peixoto_2014} as well as agglomerative \cite{blondel2008fast,ahn2010link,agglo_bayes} and recursive partitioning \cite{li2020hierarchical} procedures relying on various measures of similarity. Importantly, learning node representations characterizing structure at multiple scales of the network can facilitate network visualization and the understanding of the inner dynamics of networks. Hierarchical representation of bipartite networks is of special interest due to the fact that most unipartite hierarchical clustering algorithms do not generalize to the bipartite case beyond clustering each mode separately or transforming the bipartite network into a unipartite representation. In the work of \cite{bicl_1}, the authors used the spectral partitioning algorithm of \cite{dhillon} and then applied k-means on the spectral space to get initial bi-clusters which were followed by divisive bi-splits to create a dendrogram. In this case, the spectral embedding space was not constructed to reflect explicitly the clustering criterion. In addition to divisive procedures, agglomerative clustering has also been proposed for bipartite networks. In the work of \cite{bicl_2} a multi-objective function was designed and combined with classical community construction algorithms. One limitation here is that the network should be transformed into a unipartite structure. %Furthermore, \cite{bicl_3} used the p-value of a hypergeometric distribution measuring the probability of two entities having at least $n$ common features as the dissimilarity measure, later used for the quadratic complexity agglomerative clustering imposed.

Despite the many advantages of hierarchical structures and block models, one major limitation remains to accurately account for homophily \cite{hoff2007modeling}, which is a key characteristic of social networks. More specifically, block models have been extended to explicitly impose a community structure \cite{rosvall2007information,morup2012bayesian} but notable this only provides within-cluster homogeneity and thus homophily-like properties for the community relative to the other communities but not a hierarchy complying with such a structure. Whereas \textsc{LPCM} accounts for homophily it does not account for hierarchical structures and cluster structures are not strictly imposed beyond a prior to promoting the latent positions to form groups.
%More specifically, block models have an imposed community structure, but it is not guaranteed to obey homophily-like characteristics for intra-clusters even though they might be satisfied within clusters.

In this work, we propose a novel node representation learning approach, the Hierarchical Block Distance Model (\textsc{HBDM})\footnote{\textit{For implementation details please visit:} \href{https://github.com/Nicknakis/HBDM}{\textit{github.com/Nicknakis/HBDM}}.}, as a reconciliation between hierarchical block structures of different scales and network properties such as homophily and transitivity. In particular, we propose a framework combining embedding and hierarchical characterization for graph representation learning. Importantly, we design a hierarchical structure that respects a linearithmic total time and space complexity, in terms of the number of nodes (i.e., $\mathcal{O}(N \log N)$), and at the same time provides an accurate interpretable representation of structure at different scales. Using the \textsc{HBDM}, we embed moderate-sized and large-scale networks containing more than a million nodes and establish the performance of our model in terms of link prediction and node classification to existing prominent graph embedding approaches. We further highlight how the inferred hierarchical organization can facilitate accurate visualization of network structure even when using only $D=2$ dimensional representations providing favorable performance in all the considered GRL tasks; link prediction, node classification, and network reconstruction. Additionally, we show how our proposed framework extends the hierarchical multi-resolution structure to bipartite networks and provides the characterization of communities at multiple scales.

%% file: 2-methods.tex
\section{The Hierarchical Block Distance Model} \label{methods}
We first concentrate our study on undirected networks and later generalize our approach to bipartite graphs. We now provide the necessary definitions required throughout the paper.
%Let $\mathcal{G}=(V,E)$ be a graph where $N := \left| V \right|$ is the number of nodes and $Y_{N \times N}=\left[y_{i,j}\right]$ be the adjacency matrix of the graph such that $y_{i,j}=1$ if there is an edge between the nodes $v_i$ and $v_j$ and otherwise it is equal $0$ for all $ 1\leq i< j\leq N$.
Let $\mathcal{G}=(V,E)$ be a graph where $N := \left| V \right|$ is the number of nodes and $Y_{N \times N}=\left(y_{i,j}\right)\in \{0,1\}^{N\times N}$ be the adjacency matrix of the graph such that $y_{i,j}=1$ if the pair $(i,j) \in E$ otherwise it is equal $0$, for all $ 1\leq i< j\leq N$. We denote the latent representations of nodes by $\mathbf{Z}=(z_{i,d})\in\mathbb{R}^{N\times D}$ where each row vector, $\mathbf{z}_i\in \mathbb{R}^{D}$, indicates the corresponding embedding of node $i \in \mathcal{V}$ in a $D$-dimensional space.
% \begin{figure}[!t]
%     \subfloat[\label{total_proc}]{%
%       \includegraphics[scale=0.033]{figures/blocks_final.pdf}
%     }
%     \hfill
%     \subfloat[\label{level_1}]{%
%       \includegraphics[scale=0.15]{figures/l_1.pdf}
%     }
%     \hfill
% \subfloat[\label{level_2}]{%
%       \includegraphics[scale=0.16]{figures/l_2.pdf}
%     }
%     \hfill
% \subfloat[\label{level_3}]{%
%       \includegraphics[scale=0.18]{figures/l_3.pdf}
%     }
% %\hspace*{-3.9em}
%     \hfill
%     \subfloat[\label{final_level}]{%
%       \includegraphics[scale=0.19]{figures/final.pdf}
%     }
%     \caption{Schematics of the distance matrix calculation for a hierarchical structure of tree height $L=3$ and number of observations $N=64$.}
%     \label{fig:dummy}
% \end{figure}

%\subsection{Latent Space Models }
 The most natural choice for modeling homophily and transitivity can be found in the Latent Space Model (\textsc{LSM}) which defines an $\mathbb{R}^D$-dimensional latent space in which every node of the graph is characterized through the unobserved but informative node-specific variables $\{\mathbf{z}_i \in \mathbb{R}^D \}$. These variables are considered sufficient to describe and explain the underlying relationships between the nodes of the network. The probability of an edge occurring is considered conditionally independent given the unobserved latent positions. Consequently, the total probability distribution of the network can be written as:
\begin{equation}
    \label{eq:prob_adj}
    P(Y|\bm{Z},\bm{\theta})=\prod_{i< j}^Np(y_{i,j}|\mathbf{z}_i,\mathbf{z}_j,\bm{\theta}_{i,j}),
\end{equation}
 where $\bm{\theta}$ denotes any potential additional parameters, such as covariate regressors. A popular and convenient parameterization of  Equation \eqref{eq:prob_adj} for binary data is through the logistic regression model \cite{exp1,link2,KRIVITSKY2009204,doi:10.1198/016214504000001015}. In contrast, we adopt the Poisson regression model similar to \cite{doi:10.1198/016214504000001015} under a generalized linear model framework for the \textsc{LSM}. The use of a Poisson likelihood for modeling binary relationships in a network does not decrease the predictive performance nor the ability of the model to detect the network structure, as shown in \cite{6349745}. It also generalizes the analysis to integer-weighted graphs. In addition, the exchange of the {\it logit} to a {\it log} link function when transitioning from a Bernoulli to a Poisson model defines nice decoupling properties over the predictor variables in the likelihood \cite{karrer2011stochastic,herlau2014infinite}. Utilizing the Poisson Latent Distance Model (\textsc{LDM}) of the \textsc{LSM} family framework, the rate of an occurring edge depends on a distance metric between the latent positions of the two nodes. In our formulation, we consider the LDM Poisson rate with node-specific biases or random-effects \cite{doi:10.1198/016214504000001015,KRIVITSKY2009204} such that the expression for the Poisson rate becomes: 
\begin{equation}
    \lambda_{ij}=\exp\big(\gamma_i+\gamma_j- d(\mathbf{z}_i,\mathbf{z}_j)\big),
    \label{eqn:random_effect}
\end{equation}
where $\gamma_i \in \mathbb{R}$ denotes the node-specific random-effects and $d_{ij}(\cdot,\cdot)$ denotes any distance metric obeying the triangle inequality $\big\{ d_{ij}\leq d_{ik}+d_{kj},\:\forall(i,j,k) \in V^3 \big\}$. Considering variables 
% $\mathbf{z}$ 
$\{\mathbf{z}_i\}_{i\in V}$ as the latent characteristics, Equation \eqref{eqn:random_effect} shows that similar nodes will be placed closer in the latent space, yielding a high probability of an occurring edge and thus modeling homophily and satisfies network transitivity and reciprocity through the triangle inequality whereas the node-specific bias can account for degree heterogeneity. The conventional \textsc{LDM} rate utilizing a global bias, $\gamma^{g}$, corresponds to the special case in which $\gamma_i=\gamma_j=0.5\gamma^{g}$. As in \cite{exp1}, we presently adopt the Euclidean distance as the choice for the distance metric $d_{ij}(\cdot,\cdot)$.   

\subsection{Designing A Linearithmic Complexity}
Our goal is to design a Hierarchical Block Model preserving homophily and transitivity properties with a total complexity allowing for the analysis of large-scale networks. Our \textsc{HBDM}, defines the rate of a link between each network dyad $\{i,j\}\in V\times V$ based on the Euclidean distance, as shown in Equation \eqref{eqn:random_effect}. Therefore, we can define a block-alike hierarchical structure by a divisive clustering procedure over the latent variables in the Euclidean space. The total optimization cost of such a model though should have a linearithmic upper bound complexity to make large-scale analysis feasible. Introducing a number of clusters $K$ equal to the number of nodes $N$ in the \textsc{HBDM}, leads to the same log-likelihood as of the standard \textsc{LDM}, defining a sum over each ordered pair of the network, as:
\begin{align}
    \log P(Y|\bm{\Lambda})& =\sum_{i<j}\Big(y_{ij}\log(\lambda_{ij})-\lambda_{ij}\Big)\nonumber
\\ 
& =\sum_{i<j:y_{ij}=1}\log(\lambda_{ij})\;-\;\sum_{i< j}\lambda_{ij}\:,
    % \label{eq:log_likel_lsm}
\end{align}
For brevity, we presently ignore the linear scaling by dimensionality $D$ of the above log-likelihood
%by dimensionality $D$
function. Notably, the link contribution $\sum_{y_{i,j}=1}\log(\lambda_{i,j})$ is responsible for positioning "similar" nodes closer in the latent space, expressing the desired homophily.

  In addition, large networks are highly sparse \cite{barabasi2016network} with the number of edges being proportional to the number of nodes in the network. As a result, the computation of the link contribution is relatively cheap, scaling linearithmic or sub-linearithmic (as shown in supplementary). Most importantly, the link term removes rotational ambiguity between the different blocks of the hierarchy (as discussed later).
  %\begin{wrapfigure}{r}{0.4\textwidth}
  %\begin{center}
  %  \includegraphics[width=0.4\textwidth]{figures/homophily/edgevsnlogn.pdf}
  %\end{center}
  %\caption{Log-Log plot of number of network edges versus $N\log N$ where $N$ the number of vertices, for $70$ datasets of the SNAP library \cite{snapnets}.}\label{fig:edges}\end{wrapfigure}
  For these three reasons, no block structure is imposed on the calculation of the link contribution. The second term acts as the repelling force for dissimilar nodes and requires the computation of all node pairs scaling as $\mathcal{O}(N^2)$ making the evaluation of the above likelihood infeasible for large networks. By enforcing a block structure, i.e., akin to stochastic block models \cite{white1976social, holland1983stochastic,doi:10.1198/016214501753208735}, when grouping the nodes into $K$ clusters we define the rate between block $k$ and $k'$ in terms of their distance between centroids. A simple block structure without a hierarchy would lead to the following non-link expression:

% \begin{equation}
% \begin{split}
%     \sum_{i< j}\lambda_{ij}\approx&\sum_k^K\Bigg(\sum_{i,j \in C_k,i<j}e^{(\gamma_i+\gamma_j - ||\mathbf{z}_i-\mathbf{z}_j||_2)}+\sum_{i\in C_k<j\in C_{k'}}e^{(\gamma_i+\gamma_j - ||\bm{\mu}_k-\bm{\mu}_{k'}||_2)} \Bigg) \\= & \Bigg(\sum_k^K\sum_{i,j \in C_k}e^{(\gamma_i+\gamma_j - ||\mathbf{z}_i-\mathbf{z}_j||_2)}\Bigg)+\sum_{k<k'}e^{- ||\bm{\mu}_k-\bm{\mu}_{k'}||_2}\left(\sum_{i\in C_k}e^{\gamma_i}\right)\Bigg(\sum_{j\in C_{k'}}e^{\gamma_j}\Bigg),
%     \label{eq:log_likel_lsm_kmeans2}
%     \end{split}
% \end{equation}
\begin{align}
 \sum_{i< j}&\lambda_{ij}\!\approx\! \sum_{k=1}^{K}\!\Bigg(\!\!\sum_{\substack{i<j \\ i,j \in C_{k} }}\!\!\exp{\!\big(\gamma_i\!+\!\gamma_j \!-\! ||\mathbf{z}_i-\mathbf{z}_j||_2\big)}\nonumber
\\ 
& \!+\! \sum_{k^{'} > k}^K\sum_{i\in C_{k}}\sum_{j\in C_{k^{'}}}\exp{\big(\gamma_i + \gamma_j - ||\bm{\mu}_{k}-\bm{\mu}_{k'}||_2\big)}\Bigg),%\nonumber
% \\
% \!&=\! \sum_{k=1}^{K}\!\Bigg(\!\!\sum_{\substack{i<j \\ i,j \in C_{k} }}\!\!\exp{\!\big(\gamma_i\!+\!\gamma_j \!-\! ||\mathbf{z}_i-\mathbf{z}_j||_2\big)}\! +\!\! \sum_{k^{'} > k}^K\!\exp{(-||\bm{\mu}_{k}\!-\!\bm{\mu}_{k'}||_2)}\!\sum_{i\in C_{k}}\!\exp(\gamma_i)\!\sum_{j\in C_{k^{'}}}\!\!\exp(\gamma_j)\!\Bigg)
 \label{eq:log_likel_lsm_kmeans2}
\end{align}
where $\bm{\mu}_k$ denotes the $k$'th cluster centroid of the set $\bm{C}=\{C_1,\dots,C_K\}$, and has absorbed the dependency over the variables $\bm{Z}\in \mathbb{R}^{N\times D}$. More specifically, the cluster centroids $\bm{\mu}_k$ are implicit parameters defined as a function over the latent variables, as we will show later. Overall, considering the principle that connected and homophilic nodes will be placed closer in the latent space, this expression generalizes this principle by introducing a clustering procedure that obeys "cluster-homophily" and "cluster-transitivity" over the latent space. More specifically, we can assume that closely related nodes will be positioned in the same cluster while related or interconnected clusters will also be positioned close in the latent space, providing an accurate block structure schema. As opposed to the \textsc{LPCM} where clustering structures are imposed through a prior, the above formulation strictly defines the clustering structure as shared overall proximity between blocks as defined by the distances between centroids of the formed groups.
%Assuming equally sized clusters having $N/K$ nodes the first part scales $\mathcal{O}(N^2/K)$ whereas the second part scales $\mathcal{O}(K^2)$. As such, there is an undesirable inherent trade-off in which the first term reduces by $K$ but the second term increases quadratically. Thus, by setting $K=N/log(N)$ we reduce the first part to scale as $\mathcal{O}(N\log{N})$ but at the cost of the second term scaling $\mathcal{O}(N^2/log(N)^2)$ which for large networks is still prohibitive.
% \begin{figure}[!t]
% \centering
% \includegraphics[width=0.9\textwidth]{figures/blocks.pdf}
% %\caption{Schematics of the distance matrix calculation for a hierarchical structure of tree height $L=3$ and number of observations $N=64$.}
% \caption{Schematic representation of the distance matrix calculation for a hierarchical structure of the tree of height $L=3$ and for the number of observations $N=64$.}
% \label{fig:dist_mat_calc}
% \end{figure}
\begin{figure*}[!t]
\centering
 \subfloat[(i)]{{ \includegraphics[width=0.52\textwidth]{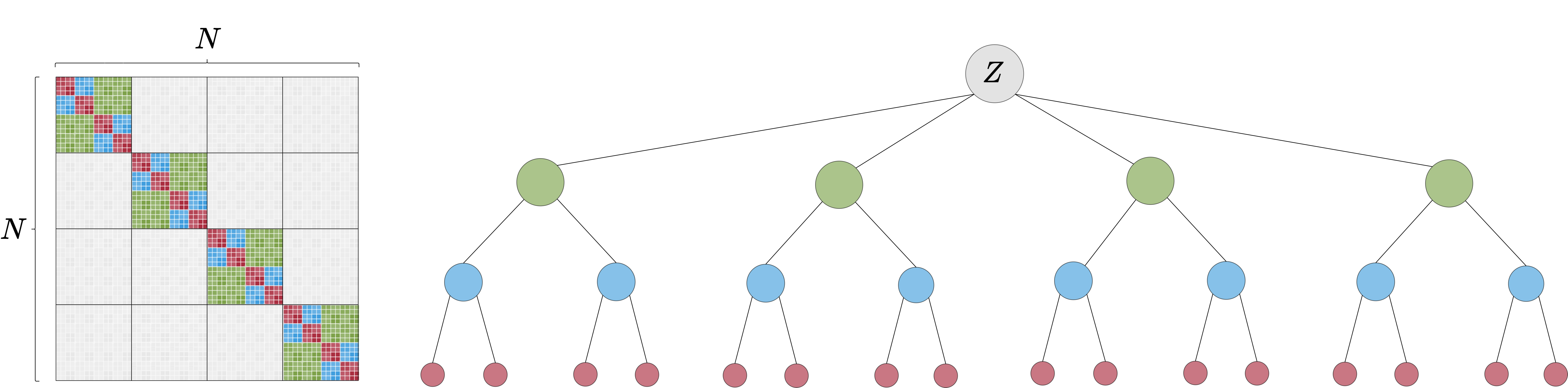} }}
\hfill
\subfloat[(ii)]{{ \includegraphics[width=0.42\textwidth]{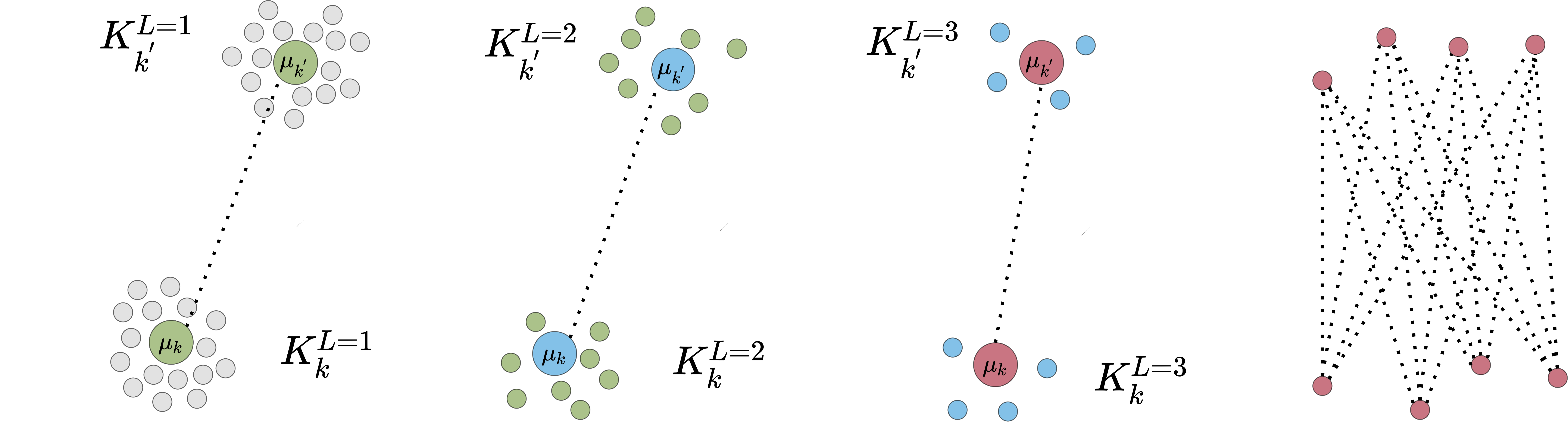} }}
%\caption{Schematics of the distance matrix calculation for a hierarchical structure of tree height $L=3$ and number of observations $N=64$.}
\caption{Schematic representation of the distance matrix calculation for a hierarchical structure of the tree of height $L=3$ and for the number of observations $N=64$.}
\label{fig:dist_mat_calc}
\end{figure*}
\subsubsection{A Hierarchical Representation}
In order to obtain the desired hierarchical representation, we define hierarchical clustering via a divisive procedure. In detail, we organize the embedded clusters into a hierarchy using a tree structure, defining a cluster dendrogram. The root of the tree is a single cluster containing the total amount of latent variable embeddings $\bm{Z}$. At every level of the tree, we perform partitioning until we obtain leaf nodes containing equal or less than the desired number of nodes, $N_{leaf}$. This number is chosen with respect to our linearithmic complexity upper bound and set as $N_{\text{leaf}}=\log N$, resulting in approximately $K=N/log(N)$ total clusters. The tree-nodes belonging to a specific tree-level are considered the clusters for that specific tree height. Every novel partition of a non-leaf node is performed only on the set of points allocated to the parent tree-node (cluster). For every level of the tree, we consider the pairwise distances of datapoints belonging to different tree-nodes as the distance between the corresponding cluster centroids, as illustrated by Fig. \ref{fig:dist_mat_calc} (ii).
%Figures \ref{level_1}, \ref{level_2} and \ref{level_3}. 
Based on these distances, we calculate the likelihood contribution of the blocks and continue with binary splits, down the tree, for the non-leaf tree-nodes. When all tree-nodes are considered as leaves, we calculate analytically the inner cluster pairwise distances for the corresponding likelihood contribution of analytical blocks, as shown in the last part of Fig. 
%\ref{final_level}. 
\ref{fig:dist_mat_calc} (ii). The latter analytical calculation comes at a linearithmic cost of $\mathcal{O}(K N_{\text{leaf}}^2)=\mathcal{O}(N \log N)$ while enforces the homophily and transitivity properties of the model since for the most similar nodes the \textsc{HBDM} behaves explicitly as the standard \textsc{LDM}.

%For the second term to be scalable, we need for $K=N/log(N)$ the contribution for the second term to scale also at $\mathcal{O}(N\log{N})$. % is proportional to the number of nodes in the network. This necessity increases the complexity of a k-means procedure as well as of computing the all-pair centroids distance matrix to $ \mathcal{O}(N^2)$. 
 %The root of the tree contains the total amount of latent variables $\bm{Z}$. At every level of the tree, we perform partitioning of the tree-nodes which are not considered as leafs. The tree-nodes belonging to a specific level are considered as the clusters for that specific tree height. Every novel partition of a non-leaf node is performed only on the set of points allocated to the parent tree-node (cluster). A node is considered as a leaf if the corresponding cluster contains less than a specific amount of datapoints which for $K=N/log(N)$ is set to be approximately equal to $N/K=\log N$. For every level of the tree, we consider the pairwise distances of datapoints belonging to different tree-nodes as the distance between the corresponding cluster centroids, as illustrated by Figures \ref{level_1}, \ref{level_2} and \ref{level_3}. Based on these distances, we calculate the likelihood contribution of these approximation blocks and continue down the tree for the non-leaf tree-nodes. In the last level (or when all tree-nodes are considered as leafs) we calculate analytically the inner cluster pairwise distances for the corresponding likelihood contribution of analytical blocks, as shown by Figure \ref{final_level}.

We can thereby define a Hierarchical Block Distance Model with Random Effects (\textsc{HBDM-Re}) as:
% \begin{eqnarray}
% %\begin{split}
%     \log P(Y|\bm{\lambda})&=\sum_{y_{i,j}=1}\Bigg(\gamma_i+\gamma_j - ||\mathbf{z}_i-\mathbf{z}_j||_2\Bigg) -\sum_{k_L}^{K_L}\Bigg(\sum_{i,j \in C_{k_L}}e^{(\gamma_i+\gamma_j - ||\mathbf{z}_i-\mathbf{z}_j||_2)}\Bigg)\\ &\qquad\qquad\qquad -\sum_{l}^{L}\left(\sum_{k_l>k_l'}e^{- ||\bm{\mu}_{k_l}-\bm{\mu}_{k_l'}||_2}\Bigg(\sum_{i\in C_{k_l}}e^{\gamma_i}\Bigg)\Bigg(\sum_{j\in C_{k_l'}}e^{\gamma_j}\Bigg)\right),
%     \label{eq:log_likel_lsm}
% %\end{split}
% \end{eqnarray}
\begin{align}
\log& P(Y|\mathbf{Z}, \bm{\gamma}) = \sum_{\substack{i < j \\ y_{i,j}=1}}\Bigg(\gamma_i+\gamma_j - ||\mathbf{z}_i-\mathbf{z}_j||_2\Bigg)\nonumber\\ & -\sum_{k=1}^{K_L}\Bigg(\sum_{\substack{i<j \\ i,j \in C^{(L)}_{k} }}\exp{(\gamma_i+\gamma_j - ||\mathbf{z}_i-\mathbf{z}_j||_2)}\Bigg)\nonumber\\ & -\sum_{l=1}^{L}\sum_{k=1}^{K_l}\sum_{k^{'}>k}^{K_l}\Bigg(\exp{(- ||\bm{\mu}^{(l)}_{k}-\bm{\mu}^{(l)}_{k'}||_2)}\nonumber\\ &\times\Big(\sum_{i\in C_{k}^{(l)}}\exp{(\gamma_i)}\Big)\Big(\sum_{j\in C_{k^{'}}^{(l)}}\exp{(\gamma_j)}\Big)\Bigg),
    \label{eq:log_likel_lsm}
\end{align}
where $l \in \{1,\ldots, L\}$ denotes the $l$'th dendrogram level, $k_l$ is the index representing the cluster id for the different tree levels, and $\bm{\mu}_{k}^{(l)}$ the corresponding centroid. We also consider a Hierarchical Block Distance Model (\textsc{HBDM}) without the random effects setting $\gamma_i=0.5\gamma^{g}$. For a multifurcating tree splitting in $K$ clusters and having $N/log(N)$ terminal nodes (clusters), the number of internal nodes are $\mathcal{O}(N/(K\log{N}))$ and each node needs to evaluate $\mathcal{O}(K^2)$ pairs providing an overall complexity of $\mathcal{O}(NK/\log{N})$, thus $K\leq\log{N}^2$ to achieve $\mathcal{O}(N\log{N})$ scaling \cite{trees}.

\subsubsection{Divisive partitioning using k-means with a Euclidean distance metric} \label{sub:centroids}
Whereas the likelihood in Equation \eqref{eq:log_likel_lsm} can be directly minimized by assigning nodes to the clusters given by the tree structure, this evaluation for all $N$ nodes scales prohibitively as $\mathcal{O}(N^2/\log{N})$. To reduce this scaling, we use a more efficient divisive partitioning procedure, minimizing the Euclidean norm $||\bm{\mu}_{k_l}-\bm{\mu}_{k_l'}||_2$. The divisive clustering procedure thus relies on the following Euclidean norm objective
\begin{equation}
    J(\mathbf{r},\bm{\mu})=\sum_{i=1}^N\sum_{k=1}^K r_{ik}||\mathbf{z}_i-\bm{\mu}_k||_2,
    \label{eqn:eucl_kmeans}
\end{equation}
where $k$ denotes the cluster id, $\mathbf{z}_i$ is the i'th data observation, $r_{ik}$ the cluster responsibility/assignment, and $\bm{\mu}_k$ the cluster centroid.

This objective function is unfortunately not accounted for by existing k-means clustering algorithms relying on the squared Euclidean norm. We therefore presently derive an optimization procedure for k-means clustering with Euclidean norm utilizing the auxiliary function framework of \cite{Tsutsu2012AnLP} developed in the context of compressed sensing. We define an auxiliary function for \eqref{eqn:eucl_kmeans} as:
\begin{equation}
     J^+(\bm{\phi},\mathbf{r},\bm{\mu})=\sum_{i=1}^N\sum_{k=1}^K r_{ik}\Bigg(\frac{||\mathbf{z}_i-\bm{\mu}_k||_2^2}{2\phi_{ik}}+\frac{1}{2}\phi_{ik}\Bigg),
    \label{eqn:eucl_kmeans_aux}
\end{equation}
where $\bm{\phi}$ are the auxiliary variables. Thereby, minimizing Equation \eqref{eqn:eucl_kmeans_aux} with respect to $\bm{\phi}_{nk}$ yields $\bm{\phi}_{ik}^*=||\mathbf{z}_i-\bm{\mu}_k||_2$ and by plugging $\bm{\phi}_{ik}^*$ back to \eqref{eqn:eucl_kmeans} we obtain $J^+(\bm{\phi}^*,\mathbf{r},\bm{\mu})=J(\mathbf{r},\bm{\mu})$ 
verifying that \eqref{eqn:eucl_kmeans_aux} is indeed a valid auxiliary function for \eqref{eqn:eucl_kmeans}. The algorithm proceeds by optimizing cluster centroids as $\bm{\mu}_k=\Big(\sum_{i\in k}\frac{\mathbf{z_i}}{\phi_{ik}}/\sum_{i\in k}\frac{1}{\phi_{ik}}\Big)$ and assigning points to centroids as $\argmin_{\bm{C}}=\sum_{k=1}^{K}\sum_{\bm{z}\in C_k}\Big(\frac{||\bm{z}-\bm{\mu}_k||_2^2}{2\bm{\phi}_{k}}+\frac{1}{2}\bm{\phi}_{k}\Big)$ upon which $\bm{\phi}_k$ is updated. The overall complexity of this procedure is $\mathcal{O}(TKND)$ \cite{Hartigan1979KMeans} where $T$ is the number of iterations required to converge. As shown in \cite{Tsutsu2012AnLP}, Equation \eqref{eqn:eucl_kmeans_aux} is a special case of a general algorithm for an $l_p(0<p<2)$ norm minimization using an auxiliary function with the algorithm converging faster the smaller $p$ is. For a detailed study of the efficiency of the optimization procedure under such an auxiliary function, see \cite{Tsutsu2012AnLP}.

A simple approach to construct the tree structure would be to use the above Euclidean k-means procedure to split the nodes into $K=N/\log(N)$ clusters and construct the tree according to agglomeration as in hierarchical clustering. Unfortunately, such a strategy is computationally prohibitive. For that, in the coarser level (first layer of the tree), we choose to split to the maximally allowed clusters of $K=\log{N}$ allowing scaling of $\mathcal{O}(N\log{N})$. It would be tempting to continue splitting into $\log{N}$ clusters, however, for a balanced multifurcating tree with $N /\log{N}$ leaf clusters, it 
will result in a height scaling as $\mathcal{O}(\log{N}/\log{\log{N}})$ and thus an overall complexity of $\mathcal{O}(N\log^{2}(N)/\log{\log{N}})$ \cite{trees}. Whereas a balanced binary tree at all levels below the root results in a height scaling as $\mathcal{O}(\log{N})$ providing an overall complexity when including the linear scaling by dimensionality $D$ of $\mathcal{O}(DN\log{N})$ (as each level of the tree defines $\mathcal{O}(DN)$ operations). Fig. \ref{fig:dist_mat_calc} (i),
illustrates the resulting tree\footnote{ For visualization purposes only, we show equally sized clusters.} for a small problem of $N=64$ nodes in which we first split into 4 ($\approx \log(64)$) clusters and subsequently create binary splits until each leaf cluster contains 4 ($\approx \log(64)$) nodes.

\subsubsection{Expressing Homophily and Transitivity}\label{hom_and_trans}
\begin{figure}[!b]
\begin{center}
\begin{minipage}[]{.38\textwidth}
\begin{center}
\subfloat[(i) Non-optimal rotation over leaf clusters.]{{ \includegraphics[width=.95\textwidth]{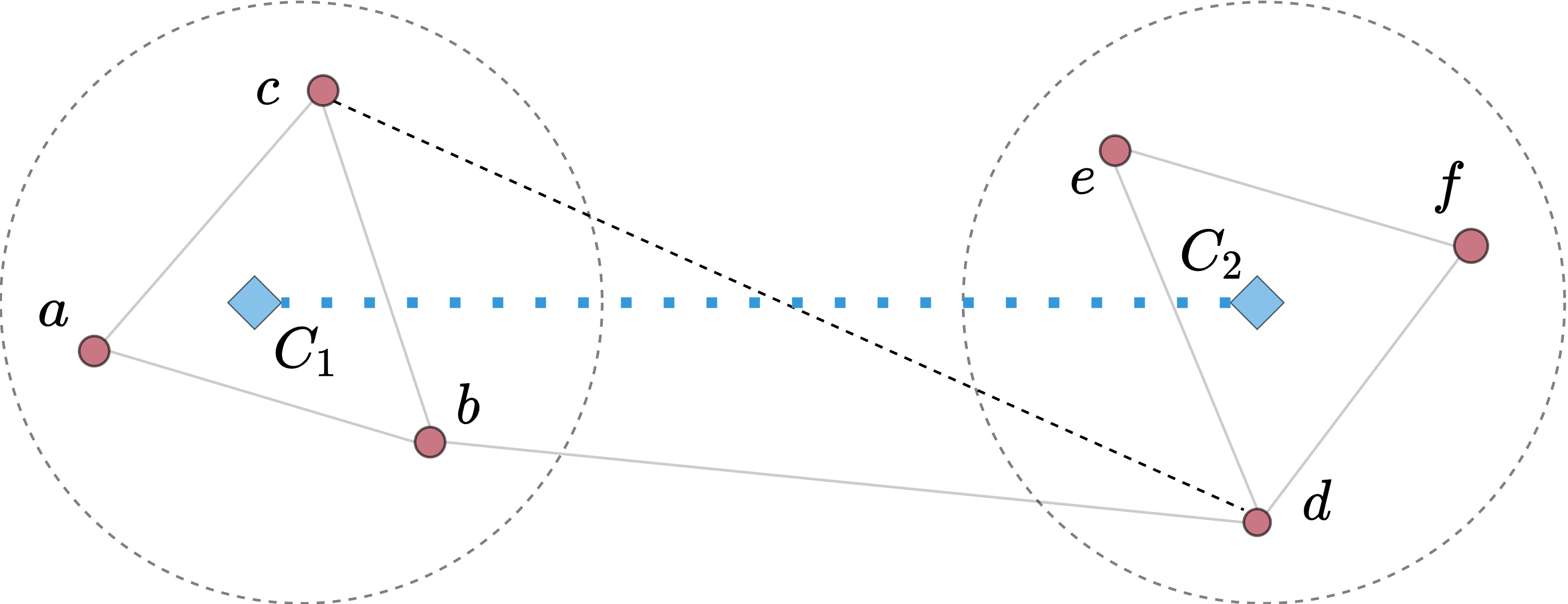} }}%
\vfill
\subfloat[(ii) Optimal rotation over leaf clusters.]{{ \includegraphics[width=.95\textwidth]{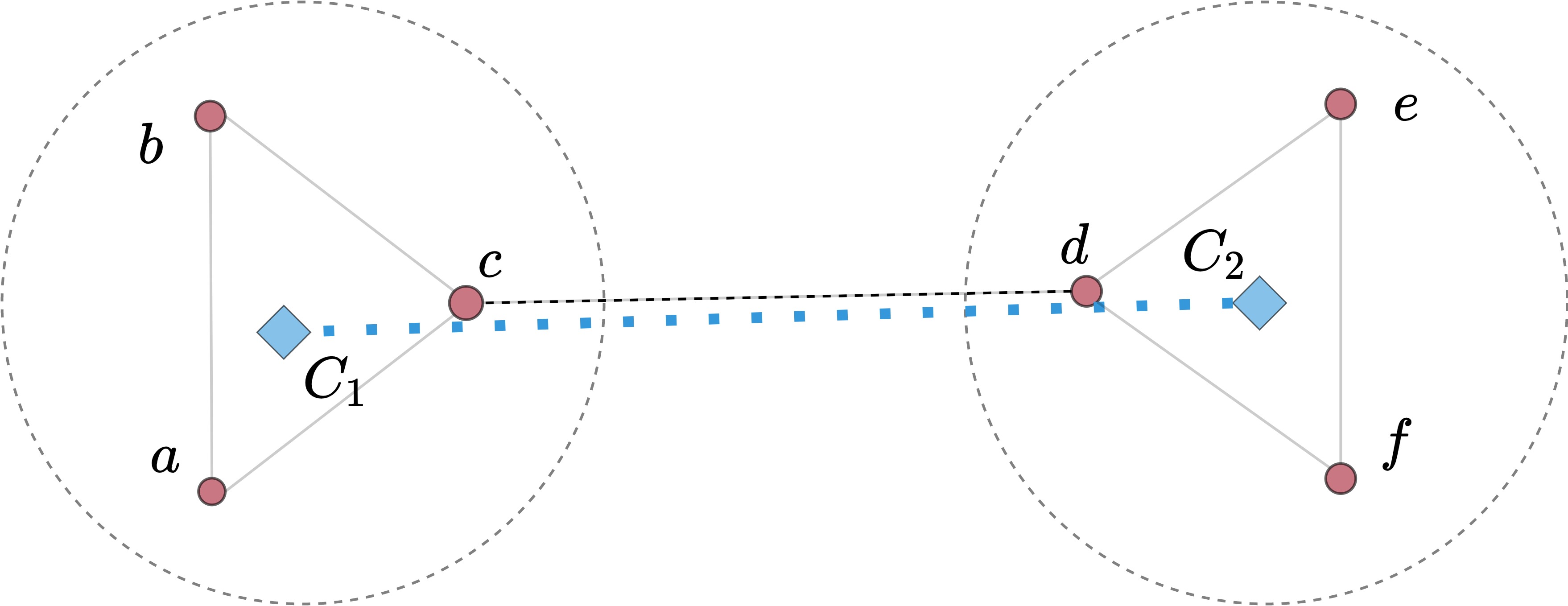} }}%
\end{center}
\end{minipage}
\hfill
\begin{minipage}[]{.48\textwidth}
\begin{center}
\subfloat[(iii) Three latent block structures.]{{ \includegraphics[width=.95\textwidth]{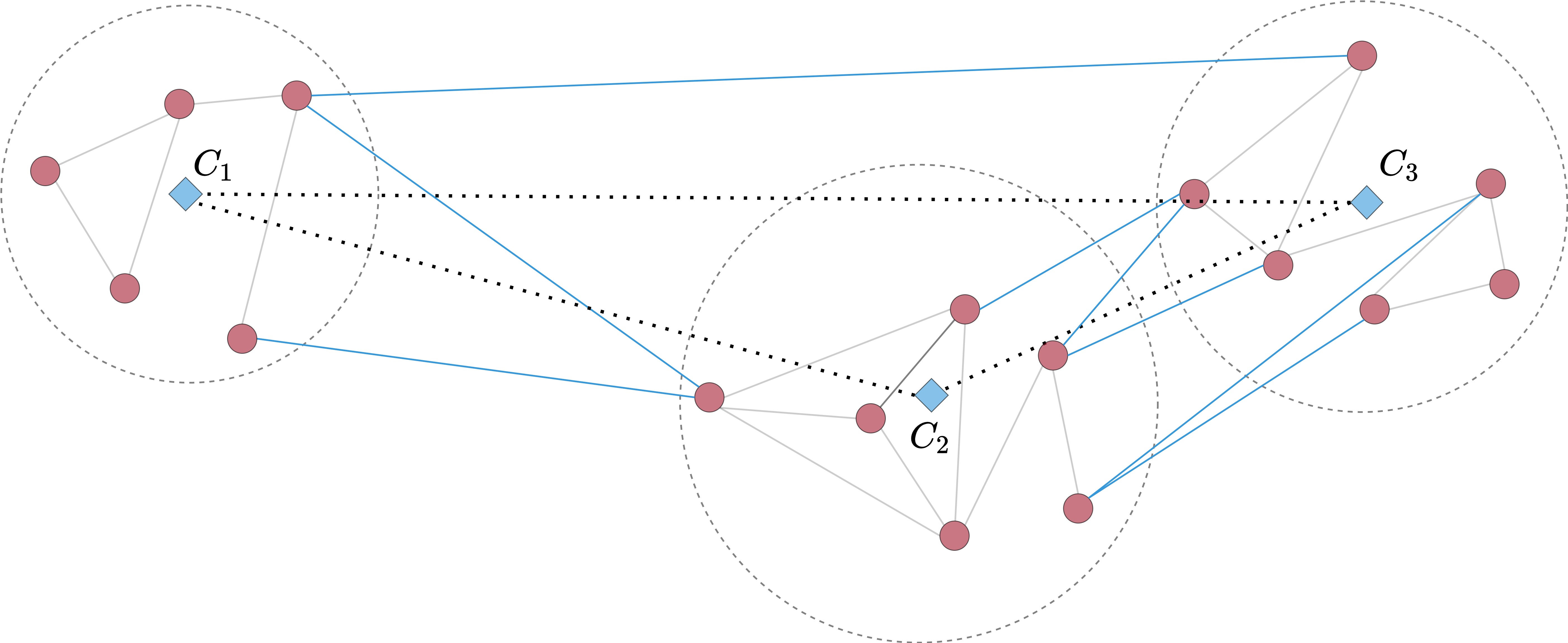} }}%
\end{center}
\end{minipage}
\end{center}
\caption{The clusters within the dashed circles denote the leaf block structures. The red circles and blue rhombuses indicate the node embeddings and the centroids, respectively. Gray lines represent the links and the dashed lines the distance between the cluster centers.}\label{fig:cluster_homophily}
%     \caption{(i) Two leaf block structures within dashed ellipses denote the clusters, and the red circles indicate the nodes in the network and the blue rhombuses the centroids. Black lines represent links in the latent space and the dashed line the distance between the cluster centers.
%   (ii) \textsc{HBDM} example of latent space where black ellipses denoting the cluster, black lines the intra-cluster links the blue lines inter-cluster links and dashed lines centroid distances.}\label{fig:cluster_homophily}
\end{figure}
A central component to preserve homophily and transitivity of the HBDM is not to approximate the link terms at the level of the block as in (hierarchical) SBMs but to calculate analytically the link contribution of the log-likelihood across the total hierarchy beyond the leaf/analytical blocks. In Fig., \ref{fig:cluster_homophily} (i) and (ii), two leaf clusters are illustrated and connected with a link. Assume that we only calculate the distance inside the blocks analytically and that both the link and non-link contributions of pairs belonging to different clusters are approximated based on their centroids' distance. This essentially would allow for any rotation of each cluster for all clusters in the hierarchy since the inner-block distances (analytical), as well as the centroid distances, would not change by such rotations, yielding exactly the same likelihood (block-level rotational invariance). In that case, homophily would be violated as, e.g., the distance between nodes $c$ and $d$ would not necessarily be smaller than the distance with other inter-cluster pairs (ex: Fig. \ref{fig:cluster_homophily} (i)), showing that the rotation of the blocks substantially impacts the homophily properties of the $\textsc{HBDM}$.
 Calculating the link contributions between the different clusters analytically solves this ambiguity since the likelihood is penalized higher when nodes $c$ and $d$ are positioned in a non-rotational-aware way. The computational cost imposed by accounting for all the link terms analytically is that the model complexity depends on the number of edges of the network (a total block structure would strictly be linearithmic in complexity). Nevertheless, we show empirically in the supplementary that the number of network edges $E$ scales linearly with $N\log N$ and thus this analytical term respects our complexity bound. In Fig. \ref{fig:cluster_homophily} (iii), we present clusters defining cases of block inter-connections of sparsely connected blocks (\{$C_1,C_3$\}, \{$C_1,C_2$\}) and densely connected blocks  \{$C_2,C_3$\}. Whereas the analytical inter-cluster links (blue lines) are responsible for fixing the block rotation the inter-cluster links also drive the cluster-level proximities of centroids %such that the triangular inequality expresses proximity at the block level, 
 ensuring cluster homophily and transitivity.

Pairwise distances in the HBDM stays invariant to rotation, reflection, and translation of the latent space due to its LDM inheritance \cite{exp1} these isometries can be resolved via a Singular-Value-Decomposition procedure as provided in the supplementary.
%Note that we have unique node representations up to isometries (reflection, translation and rotation) on Euclidean spaces \cite{exp1} for given random effect terms.  Since the link and non-link contributions between clusters in the objective function are approximated based on their centroids distances, we can also apply local rotation operations on the embeddings of a cluster without changing the loss function value. In the following lemma, we address this problem by showing that an edge from a cluster to an external node reduces the cluster's degree of freedom for rotation by one. 
Whereas the analytical link term calculations provide rotational awareness to the \textsc{HBDM} clusters, we continue by investigating the conditions in which a continuous operation defining infinitesimal rotations (with respect to the cluster centroid) is admissible leaving the loss function of Equation \eqref{eq:log_likel_lsm} invariant to continuous rotations. % since \textsc{HBDM} calculates the non-link contributions between clusters in the objective function based on their centroids distances. 
In Lemma \ref{lemma} (proof given in the supplementary material), we start our investigation of this problem by showing that blocks with a unique inter-cluster link connection reduce the clusters' degree of rotational freedom by one. 

\begin{lemma}\label{lemma}
 Let $\mathcal{G}=(\mathcal{V}, \mathcal{E})$ be a graph and let $\mathcal{C}$ be a cluster with its centroid located at $\boldsymbol{\mu}\in \mathbb{R}^D$ having an edge $(i,j)\in \mathcal{E}$ for some $i\in \mathcal{C}$ and $j\in \mathcal{V}\backslash\mathcal{C}$ such that $\mathbf{z}_i \neq \boldsymbol{\mu}$. If $\mathbf{\tilde{z}}_i=\boldsymbol{\mu}+\mathbf{R}(\boldsymbol{\theta})(\mathbf{z}_i-\boldsymbol{\mu})$ such that $\mathbf{R}(\boldsymbol{\theta})$ is a rotation matrix acting on the embeddings of nodes in cluster $\mathcal{C}$, then the maximum degree of freedom of any infinitesimal $\lambda_{ij}$-invariant rotation is defined by $\boldsymbol{\theta}\in\mathbb{R}^{D-2}$.
 \end{lemma}
A direct consequence of Lemma \ref{lemma} is that for a two-dimensional embedding, there is no possible continuous rotation of a cluster having only one external edge. Since there is a path from one node to all others in a connected graph, every cluster must have at least one external link. For the general case of blocks having multiple inter-cluster edges, rotations preserving the total sum of pairwise distances among node embeddings are highly unlikely, as discussed in the supplementary. Consequently, we can for connected networks expect uniqueness of a (local) minima solutions with no continuous admissible rotations leaving the HBDM loss function of Equation \eqref{eq:log_likel_lsm} invariant.

\begin{table}[!hb]
\caption{Complexity analysis of methods. $N := \left| V \right|$ is the vertex set, $\left| E\right|$: edge set, $\mathcal{W}$: number of walks, $\mathcal{L}$: walk length, $H$: height of the hierarchical tree, $D$: node representation size, $k$: number of negative instances, $q$: order value, $c$: Chebyshev expansion order, $\gamma$: window size, $\alpha_1$ and $\alpha_2$ constants such as $\alpha_1,\alpha_2\ll N$.} 
\label{tab:complexities}
\begin{center}
\resizebox{0.31\textwidth}{!}{%
\begin{tabular}{rc}\toprule
Method & Complexity\\\cmidrule(rl){1-1}\cmidrule(rl){2-2}
\textsc{DeepWalk} & $\mathcal{O}\left(\gamma N \log{(N)} \mathcal{W} \mathcal{L} \mathcal{D}  \right)$ \\
\textsc{Node2Vec} & $\mathcal{O}\left(\gamma N \mathcal{W} \mathcal{L} \mathcal{D}k \right)$  \\
\textsc{LINE} & $\mathcal{O}\left( |E| D k \right)$ \\
\textsc{NetMF} & $\mathcal{O}\left(N^2 D\right)$  \\
\textsc{NetSMF} & $\mathcal{O}\left( |E|(\gamma+D)+ND^2 + D^3 \right)$ \\
\textsc{RandNE} & $\mathcal{O}\left(ND^2 + |E|Dq\right)$ \\
\textsc{LouvainNE} & $\mathcal{O}\left( |E| \mathcal{H} + N D \right)$ \\
\textsc{ProNE} & $\mathcal{O}\left(ND^2 + |E|c\right)$ \\
\textsc{Verse} & $\mathcal{O}\left(N(\mathcal{W} + kD)\right)$ \\
\textsc{HBDM} & $\mathcal{O}\left(\alpha_2N\log{(N)} D\right)$ \\\bottomrule
 & 
\end{tabular}%
}
\end{center}

\end{table}

\subsubsection{Extension to Bipartite Networks}
Our proposed frameworks, \textsc{HBDM} and \textsc{HBDM-Re} generalize to both directed and bipartite graphs. In the following, we provide the mathematical extension for the bipartite case (the directed network formulation of our proposed model can be considered a special case of the bipartite framework in which self-links are removed and thus omitted from the below log-likelihood).
%\todo{Give the Filmtrust details} 
For a bipartite network with adjacency matrix $Y^{N_1 \times N_2}$ we can formulate the log-likelihood as:
% \begin{eqnarray}
% %\begin{split}
%     \log P(Y|\bm{\lambda})&=\sum_{y_{i,j}=1}\Bigg(\psi_i+\omega_j - ||\mathbf{w}_i-\mathbf{v}_j||_2\Bigg) -\sum_{k_L}^{K_L}\Bigg(\sum_{i,j \in C_{k_L}}e^{(\psi_i+\omega_j - ||\mathbf{w}_i-\mathbf{v}_j||_2)}\Bigg)\\ &\qquad\qquad\qquad -\sum_{l}^{L}\left(\sum_{k_l,k_l'}e^{- ||\bm{\mu}_{k_l}-\bm{\mu}_{k_l'}||_2}\Bigg(\sum_{i\in C_{k_l}}e^{\psi_i}\Bigg)\Bigg(\sum_{j\in C_{k_l'}}e^{\omega_j}\Bigg)\right),
%     \label{eq:log_likel_lsm_bip}
% %\end{split}
% \end{eqnarray}
\begin{align}
\log& P(Y|\bm{\Lambda})=\sum_{\substack{i < j \\ y_{i,j}=1}}\Bigg(\psi_i+\omega_j- ||\mathbf{w}_i-\mathbf{v}_j||_2\Bigg) \nonumber\\ &-\sum_{k_L=1}^{K_L}\Bigg(\sum_{i,j \in C_{k_L}}\exp{(\psi_i+\omega_j - ||\mathbf{w}_i-\mathbf{v}_j||_2)}\Bigg)\nonumber
\\ 
& -\sum_{l=1}^{L}\sum_{k=1}^{K_l}\sum_{k^{'} > k}^{K_l}\Bigg(\exp{(- ||\bm{\mu}_{k}^{(l)}-\bm{\mu}_{k'}^{(l)}||_2)}\nonumber\\ &\times\Big(\sum_{i\in C_{k}^{(l)}}\exp{(\psi_i)}\Big)\Big(\sum_{j\in C_{k^{'}}^{(l)}}\exp{(\omega_j)}\Big)\Bigg),\label{eq:log_likel_lsm_bip}
\end{align}
where $\{\bm{\mu}_k^{(l)}\}_{k=1}^{K_L}$ are the latent centroids which have absorbed the dependency of both sets of latent variables $\{\mathbf{w}_i,\mathbf{v}_j\}$ while we define the Poisson rate as:
\begin{equation}
    \lambda_{ij}=\exp\big(\psi_i+\omega_j- d(\mathbf{w}_i,\mathbf{v}_j)\big),
    \label{eqn:random_effect_bip}
\end{equation}
where $\psi_i$ and $\omega_j$ are the corresponding random effects and $\{\mathbf{w}_i$, $\mathbf{v}_j\}$ are the latent variables of the two disjoint sets of the vertex set of sizes $N_1$ and $N_2$, respectively. In this setting, we use our divisive Euclidean distance hierarchical clustering procedure over the concatenation $\mathbf{z}=[\mathbf{w};\mathbf{v}]$ of the two sets of latent variables. Therefore, we define an accurate hierarchical block structure for bipartite networks, with each block including nodes from both of the two disjoint modes. Here, a centroid is considered a leaf if the corresponding tree-cluster contains less than $\log (N_1)$ of the latent variables $\{\mathbf{w}_i\}_{i=1}^{N_1}$ or less than $\log(N_2)$ of $\{\mathbf{v}\}_{j=1}^{N_2}$.

\subsubsection{Complexity Comparison}
TABLE \ref{tab:complexities} provides a comparison between time complexities of several prominent GRL methods in terms of their Big $\mathcal{O}$ notation, similar to \cite{verse}. We observe that our proposed \textsc{HBDM} is positioned as one of the most competitive frameworks. In terms of space complexity, our model defines a linearithmic complexity contrary to the majority of the considered baselines which are usually characterized by a quadratic space complexity \cite{verse}. (For a more detailed discussion please visit the supplementary.)

%% file: 4-experiments.tex
\section{Experiments}

\begin{table*}[!t]
\begin{center}
\caption{Statistics of undirected networks. $N$: number of nodes, $|E|$: number of edges.}
\label{tab:network_statistics}
\resizebox{0.72\textwidth}{!}{%
% \begin{tabular}{rccccccc}\toprule
%  & \textsl{AstroPh}\cite{astroph_grqc_hepth} & \textsl{GrQc}\cite{astroph_grqc_hepth} & \textsl{Facebook}\cite{facebook} & \textsl{HepTh}\cite{astroph_grqc_hepth} & \textsl{YouTube}\cite{youtube} & \textsl{Flickr} \cite{youtube_flickr} & \textsl{Flixster}\cite{flixster} \\\midrule
 \begin{tabular}{rcccccccccc}\toprule
 & \textsl{Cora} & \textsl{Dblp} & \textsl{AstroPh} & \textsl{GrQc} & \textsl{Facebook} & \textsl{HepTh} & \textsl{Amazon} & \textsl{YouTube} & \textsl{Flickr} & \textsl{Flixster} \\\midrule
$N$ & 2,708 & 27,199 & 17,903 & 5,242 & 4,039 & 8,638 & 334,868 & 1,138,499 & 1,715,255 & 2,523,386 \\
$|E|$ & 5,278 & 66,832 & 197,031 & 14,496 & 88,234 & 24,827 & 925,876 & 2,990,443 & 15,555,042 & 7,918,801 \\\bottomrule
%$d^*$ & 22.010 & 5.531 & 43.691 & 5.748 & 5.253 & 18.137 & 6.276 \\\bottomrule
\end{tabular}%
}
\end{center}
\end{table*}
\label{experiments}
    We extensively evaluate the performance of our method compared to baseline graph representation learning approaches on networks of various sizes and structures.  We have conducted all the experiments regarding the \textsc{HBDM} and \textsc{HBDM-Re} on a $32$ GB Tesla V100 GPU machine with $5120$ CUDA cores, and a $1380$ MHz clock. For the \textsc{HBDM} and \textsc{HBDM-Re} models, we optimize the negative log-likelihood via the Adam \cite{kingma2017adam} optimizer with learning rate $lr \in [0.01,0.1]$. For both frameworks, we build the hierarchical structure by running the k-means procedure every $t=25$ iterations. Experiments regarding the baselines have been conducted on an Intel Xeon Gold 6342 CPU with 24 cores, 2800 MHz clock, and $512$ GB memory. The implementation for \textsc{HBDM} and \textsc{HBDM-Re} is GPU-focused using PyTorch $1.12.1$, exploiting parallel computations (running the frameworks on a CPU machine leads to substantially higher runtimes). We argue, that runtime comparison in terms of real-time is a biased estimate between different models since it correlates highly with the programming language, parallelization schemes, etc. For that, we instead compare theoretical complexities in terms of their Big $\mathcal{O}$ notation. In all TABLES, we denote with bold digits the best-performing score while we underline the second-best. %for all networks. 
\textbf{Datasets:} We have performed the experiments on ten undirected networks of various sizes and structures: a citation network (\textsl{Cora} \cite{cora}), social interaction graphs (\textsl{Facebook} \cite{facebook}, \textsl{YouTube} \cite{amazon_youtube, youtube_flickr},  \textsl{Flickr} \cite{youtube_flickr}, \textsl{Flixster} \cite{flixster}), product-label network (\textsl{Amazon} \cite{amazon_youtube}) and collaboration networks (\textsl{Dblp} \cite{dblp}, \textsl{AstroPh} \cite{astroph_grqc_hepth}, \textsl{GrQc} \cite{astroph_grqc_hepth}, \textsl{HepTh} \cite{astroph_grqc_hepth}). Each network is considered as unweighted for the consistency of the experiments. The detailed statistics of the networks are provided by TABLE \ref{tab:network_statistics}. All of the considered networks have been widely adopted and extensively used as benchmarks in the GRL literature \cite{snapnets}.

\textbf{Baseline Methods:} In our experiments, we have run various graph representation learning methods in order to evaluate the performance of our approach. The prominent GRL frameworks used in this study are: (i) \textsc{DeepWalk} \cite{deepwalk-perozzi14}, (ii) \textsc{Node2Vec} \cite{node2vec-kdd16}, (iii) \textsc{LINE} \cite{line}, (iv) \textsc{NetMF} \cite{netmf-wsdm18}. In addition, we consider five scalable graph embedding approaches: (v) \textsc{NetSMF} \cite{netsmf-www2019}, (vi) \textsc{RandNE} \cite{randne-icdm18}, (vii) \textsc{ProNE} \cite{prone-ijai19}, (viii) \textsc{LouvainNE} \cite{louvainNE-wsdm20}, (ix) \textsc{Verse} \cite{verse}. For more details see the supplementary material. In our analysis, we considered \textsc{GraphSage} \cite{graphsage_hamilton} as a prominent member of the family of Graph Neural Networks (GNNs). Our study focuses on the setting where node meta-data are not available. In such a setting, \textsc{GraphSage} was characterized by a close-to-random performance and thus not presented.

\subsection{Link Prediction}
%We report results for the area under the curve of the receiver operator characteristic (AUC) whereas we defer the corresponding Precision-Recall AUC to the supplementary material. For the experimental setup, we follow the commonly applied strategy \cite{deepwalk-perozzi14, node2vec-kdd16}, and we remove half of the edges of a given network by keeping the residual network connected. Since this strategy is not feasible for large-scale networks, we hide 30\% of the edges for these networks.  The edges in the residual network form the positive samples for the training set, and the removed links are considered the positive instances for the testing set. We sample the same number of node pairs that are not the edges of the original network to construct the negative instances for the testing set. We utilize the residual network to learn the node embeddings and design a feature vector for each node pair sample by applying a binary operator \cite{node2vec-kdd16}. The detailed list of the operators are given in supplementary. %The detailed list of the operators are given in Table \ref{tab:binary_operators}. 
%For \textsc{HBDM} and \textsc{HBDM-Re} the predictions are made directly based on the learned Poisson rates of the test set pairs $\{ij\}$, i.e. $\lambda_{ij}=\exp{(\gamma_i+\gamma_j-||\bm{z}_i-\bm{z}_j||_2)}$. Error bars for the following AUC scores were found to be in the scale of $10^{-3}$ and thus provided in the supplementary material.

We report results for the area under the curve of the receiver operator characteristic (AUC). For the experimental setup, we follow the commonly applied strategy \cite{deepwalk-perozzi14, node2vec-kdd16} and remove half of the edges while keeping the residual network connected. This strategy is not feasible for large-scale networks since checking if the residual network stays connected after each removed link results in a high runtime complexity. For that, we hide 30\% of the edges for these networks and extract the giant component (after the link removal) which is treated as the residual network. Extensive details for the link prediction experiments, as well as, Precision-Recall AUC scores are given in the supplementary. Error bars across $5$ re-runs for the following AUC scores were found to be on the scale of $10^{-3}$ and thus negligible.

\begin{figure*}
  \centering
 \subfloat[(a)]{{ \includegraphics[width=0.25\textwidth]{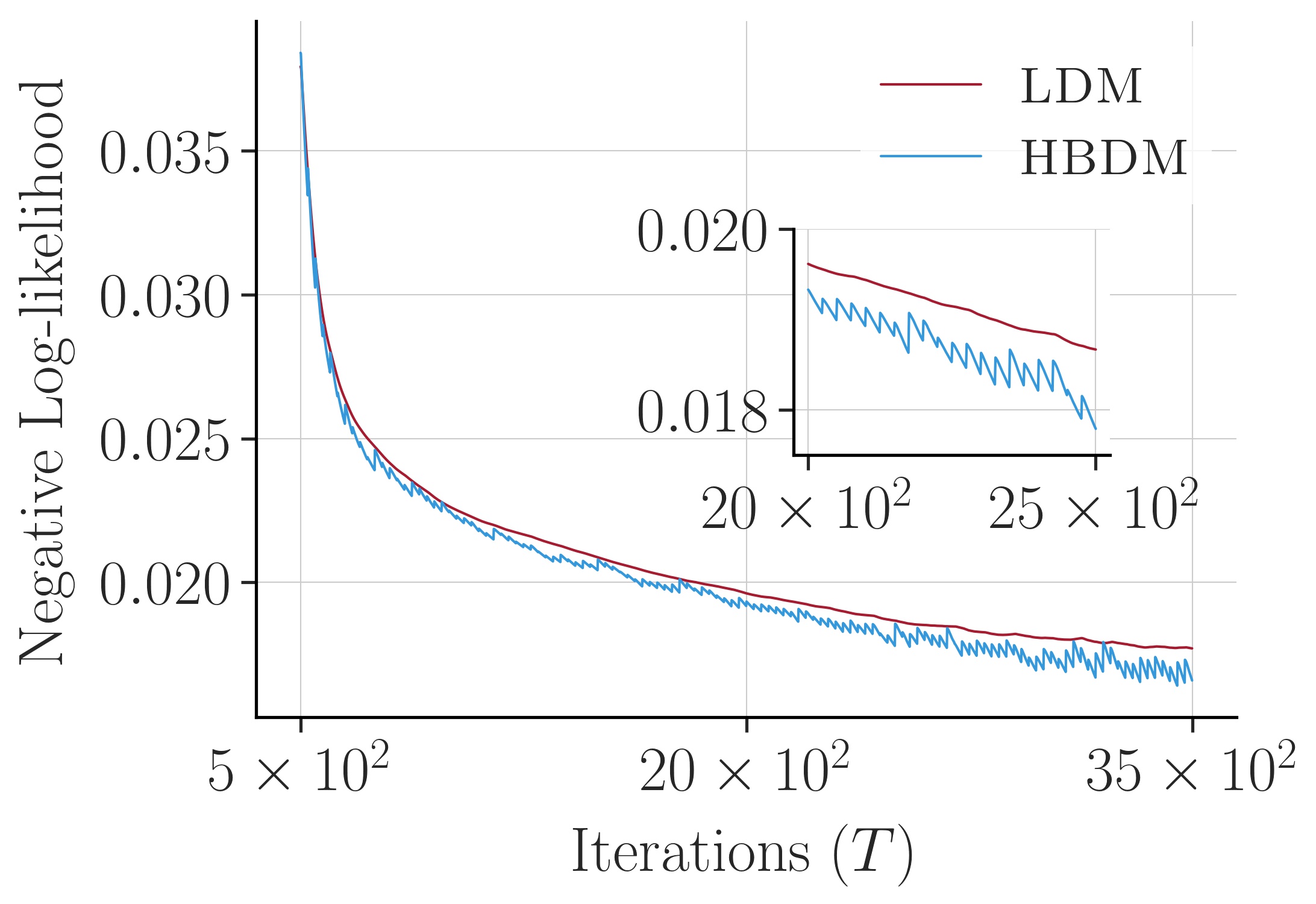}\label{fig:nll_comparison} }}
  \hfill
  \subfloat[(b)]{{ \includegraphics[width=0.25\textwidth]{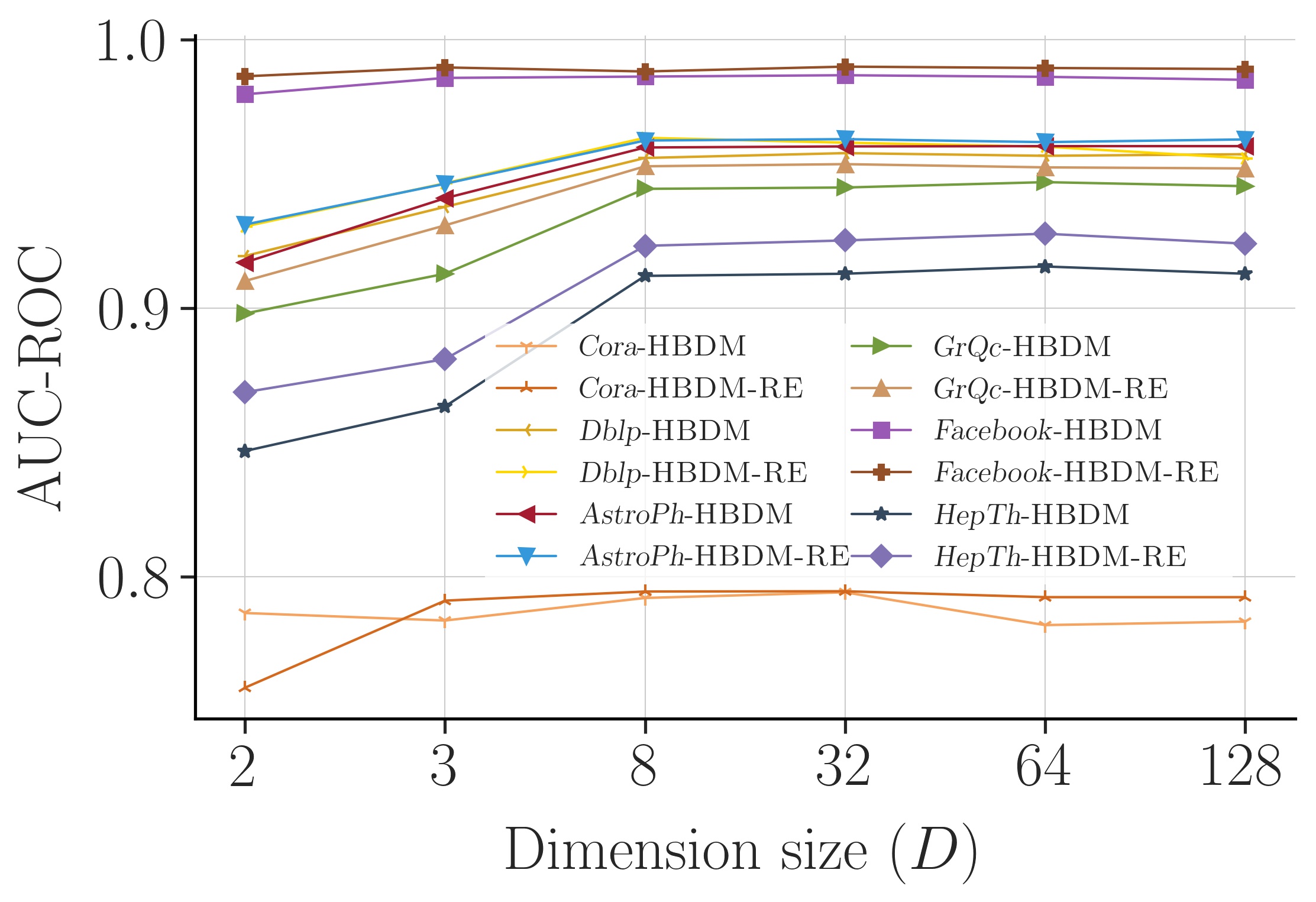}\label{fig:auc_vs_dim} }}\hfill
  \subfloat[(c)]{{ \includegraphics[width=0.2\textwidth]{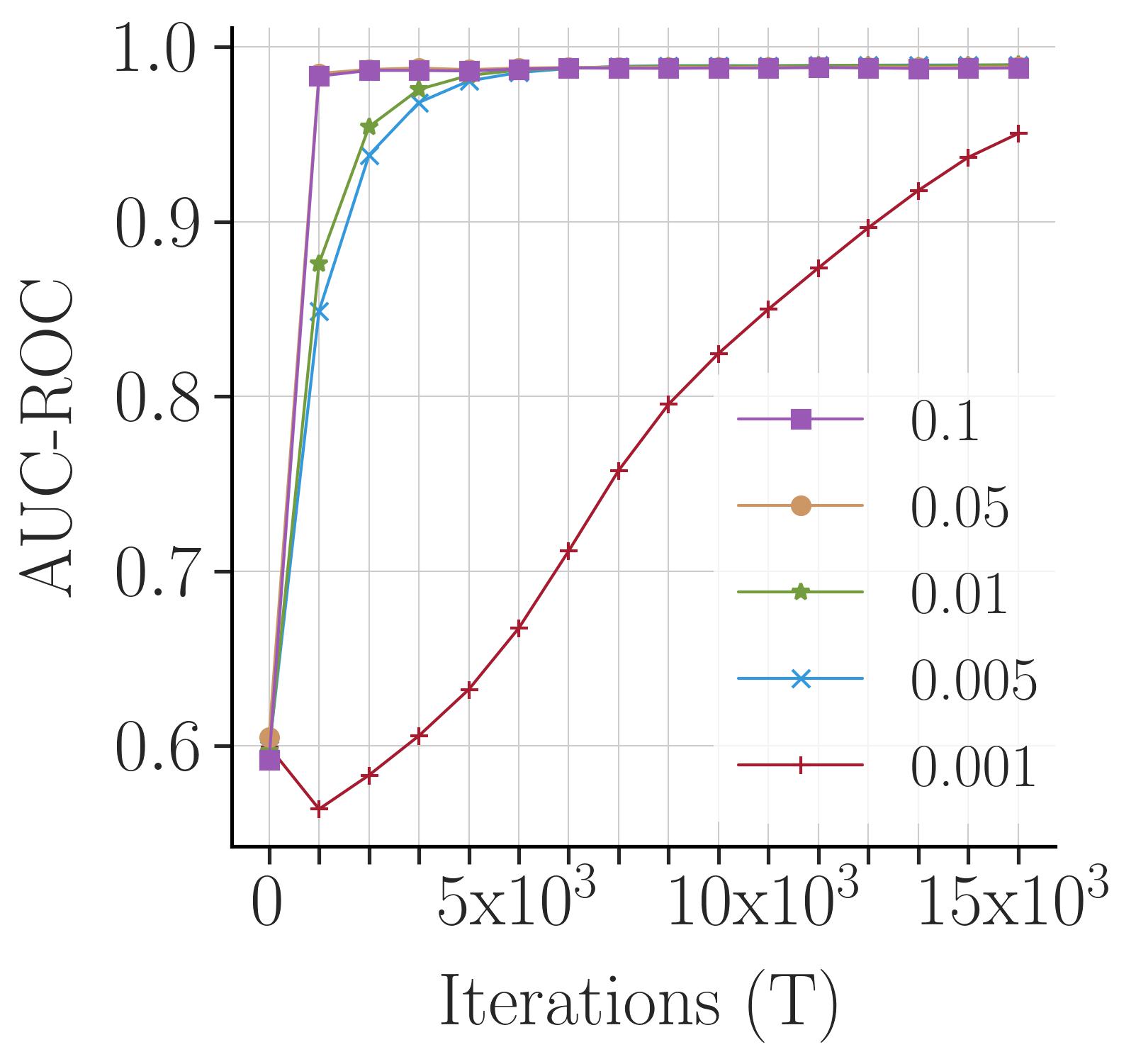}\label{fig:auc_vs_lr} }}\hfill
  \subfloat[(d)]{{ \includegraphics[width=0.2\textwidth]{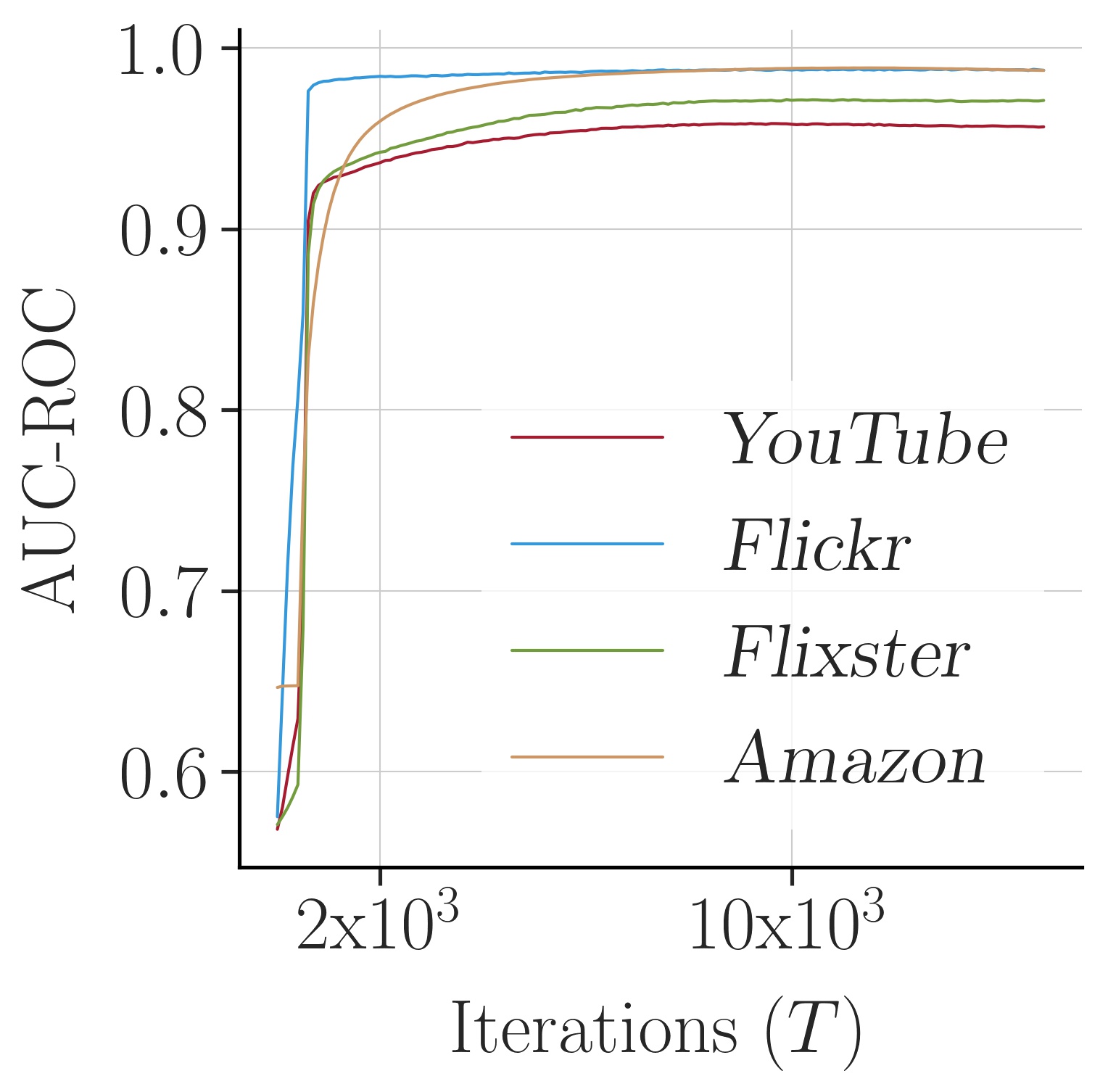}\label{fig:roc_iteration} }}
  \caption{(a) NLL comparison between \textsc{HBDM} and \textsc{LDM} for \textsl{Facebook} with $D=2$. (b) AUC-ROC performance over various networks for varying embedding sizes. (c) Performance sensitivity over different learning rate choices for the optimizer in terms of AUC-ROC scores for the \textsl{Facebook} network. (d) AUC scores of \textsc{HBDM-Re} in terms of iterations sensitivity for large-scale networks.}%\label{fig:roc_d} I removed this captions to detect the problems in referencing the figures in text. Some references are wrong because of the split
\end{figure*}

\textbf{Effectiveness and Efficiency of the Multi-Scale Approximation}: In Fig.~\ref{fig:nll_comparison}, we provide an effectiveness analysis of the \textsc{HBDM} likelihood when contrasted with its full likelihood estimation evaluated on the moderate-sized network of Facebook  (results for more networks are provided in the supplementary material). We here observe that the \textsc{HBDM} likelihood essentially approximates the true full likelihood providing systematically slightly lower likelihood estimates which we attribute to the small structural differences between calculating the distances analytically versus in a hierarchical block manner. A close approximation to the true likelihood provides evidence for multi-scale structures that characterize networks, yielding a high effectiveness of the \textsc{HBDM} framework. In addition, in Fig.~\ref{fig:nll_comparison} we see fluctuations in the likelihood which is an immediate result of building the network hierarchy from scratch every $25$'th iteration. Importantly, despite the fact that k-means is notoriously known to be an NP-hard problem \cite{kmeans_NP1,kmeans_NP2}, we observe that rebuilding the hierarchy has a minimum effect on the value of the likelihood, highlighting the stability of the inferred hierarchy in the \textsc{HBDM}. Furthermore, TABLE \ref{tab:an_vs_app} conveys information about the comparison between an \textsc{HBDM} (approx) framework where all link distances are approximated by the centroid distances and the proposed \textsc{HBDM} where link distances are calculated analytically. We witness for the \textsl{Facebook} network how the rotational awareness induced by explicitly accounting for all links in the likelihood (as explained in subsection \ref{hom_and_trans}) increases the predictive capability of the model and thus its efficiency (similar results were obtained for all networks).

\textbf{Moderate-Sized Networks:} Results for the moderate-sized networks are given in TABLE~\ref{tab:auc_moderate_networks}. The symbol "-" indicates that the running time of the corresponding model takes more than 20 hours and "x" shows that the method is not able to run due to insufficient memory space. We observe that the \textsc{HBDM} and \textsc{HBDM-Re} perform significantly better or on par with the performance of the considered baseline approaches. In particular, the \textsc{HBDM} and \textsc{HBDM-Re} perform better than all the non-\textsc{LDM} baselines when $D=2$. It highlights the superiority of \textsc{LDM}s in learning very low-dimensional network representations that accurately account for the network structure. %Importantly, embeddings of $D=2$ can directly be visualized without requiring additional dimensionality reduction operations and potential information loss. 
We further observe that representing degree heterogeneity with random effects provides extended representational power as the \textsc{HBDM-Re} consistently outperforms the \textsc{HBDM}. Comparing our framework with the classic \textsc{LDM-Re} and \textsc{LDM}, we mostly see on-par results experimentally which we attribute to the hierarchical structure well-preserving properties of homophily and transitivity. %In Figure~\ref{fig:nll_comparison}, we provide the analysis of the \textsc{HBDM} likelihood when contrasted with its full likelihood estimation evaluated on the moderate-sized network of Facebook  (results for more networks are provided in the supplementary material). We here observe that the \textsc{HBDM} likelihood essentially approximates the true full likelihood providing systematically slightly lower likelihood estimates which we attribute to the small structural differences between calculating the distances analytically versus in a hierarchical block manner. In addition, in Figure~\ref{fig:nll_comparison} we see fluctuations in the likelihood which is an immediate result of building the network hierarchy from scratch every $25$'th iteration. Importantly, despite the fact that k-means is notoriously known to be an NP-hard problem \cite{kmeans_NP1,kmeans_NP2}, we observe that rebuilding the hierarchy has a minimum effect on the value of the likelihood, highlighting the stability of the inferred hierarchy in the \textsc{HBDM}. 
 \begin{table}[!b]
\centering
\caption{AUC-ROC scores for varying dimension sizes on the \textsl{Facebook} network for a model approximating the link terms (top two rows) and for the proposed model which calculates analytically the link terms (bottom two rows).}
\label{tab:an_vs_app}
\resizebox{0.35\textwidth}{!}{%
\begin{tabular}{rcccccc}\toprule
Dimension (D) & $2$ & $3$ & $8$ & $32$ & $64$ & $128$ \\\cmidrule(rl){1-1}\cmidrule(rl){2-2}\cmidrule(rl){3-3}\cmidrule(rl){4-4}\cmidrule(rl){5-5}\cmidrule(rl){6-6}\cmidrule(rl){7-7}
\textsc{HBDM} (approx) & .656  & .797  & .946  &.943   &.940   & .945  \\
\textsc{HBDM-Re} (approx) & .802  & .838  & .909  &.932   &.940   & .942    \\
\cmidrule(rl){2-2}\cmidrule(rl){3-3}\cmidrule(rl){4-4}\cmidrule(rl){5-5}\cmidrule(rl){6-6}\cmidrule(rl){7-7}
\textsc{HBDM}           & .980  & .986  & .986  & .987  & .986  & .985  \\
\textsc{HBDM-Re}         & .986  & .990  & .988  & .989  & .989  & .989  \\  \bottomrule
\end{tabular}%
}
\end{table}

\begin{table}[!t]
\caption{AUC scores for representation sizes of $2$ and $8$ over moderate-sized networks.}\label{tab:auc_moderate_networks}
\begin{center}
\resizebox{0.48\textwidth}{!}{%
\begin{tabular}{rcccccccccccc}\toprule
\multicolumn{1}{r}{} &
  \multicolumn{2}{c}{\textsl{AstroPh}} &
  \multicolumn{2}{c}{\textsl{GrQc}} &
  \multicolumn{2}{c}{\textsl{Facebook}} &
  \multicolumn{2}{c}{\textsl{HepTh}}&
  \multicolumn{2}{c}{\textsl{Cora}}&
  \multicolumn{2}{c}{\textsl{DBLP}}
  \\\cmidrule(rl){2-3}\cmidrule(rl){4-5}\cmidrule(rl){6-7}\cmidrule(rl){8-9}\cmidrule(rl){10-11}\cmidrule(rl){12-13}
\multicolumn{1}{r}{Dimension ($D$)} & $2$ & $8$   & $2$ & $8$   & $2$ & $8$   & $2$ & $8$  & $2$ & $8$ & $2$ & $8$  \\\cmidrule(rl){1-1}\cmidrule(rl){2-2}\cmidrule(rl){3-3}\cmidrule(rl){4-4}\cmidrule(rl){5-5}\cmidrule(rl){6-6}\cmidrule(rl){7-7}\cmidrule(rl){8-8}\cmidrule(rl){9-9}\cmidrule(rl){10-10}\cmidrule(rl){11-11}\cmidrule(rl){12-12}\cmidrule(rl){13-13}
\textsc{DeepWalk}    & .831 & .945 & .845 & .919 & .958 &  .986 & .773 & .874 &.684 &.782 &.803 &.939 \\
\textsc{Node2Vec}    & .825 & .950 & .809 & .884 & .914 & \textbf{.988} & .780 & .881&.640 &.776 &.803 &.945 \\
\textsc{LINE} & .632 & .910 & .688 & .920 & .751 & .980 & .659 & .874 &.634 &.521 &.625 &.503 \\
\textsc{NetMF} & .800 & .814 & .830 & .860 & .872 & .935 & .757 & .792 &  .629  & .739  &.838 &.858  \\
\textsc{NetSMF}      & .828 & .891 & .756 & .805 & .907 & .976 & .705 & .810&.605 &.737 &.766 &.857  \\
\textsc{RandNE}      & .524 & .554 & .534 & .560 & .614 & .657 & .519 & .509&.508 &.556 &.508 &.517  \\
\textsc{LouvainNE}   & .798 & .813 & .861 & .868 & .957 & .958 & .774 & .874& .767 &.747 &.900 &.904  \\
\textsc{ProNE}  & .768	& .907	& .818 & .883 & .900 & .971 & .678 & .823 &.675 &.764 &.813 &.924   \\
\textsc{Verse}  & .899  &\textbf{.974} 	&.885  &.941  &.970  &\textbf{.992}  &.844 & .910 &.749  &.760 & .910 &.955   \\
\textsc{LDM} & {\ul .925} & x & .915 & .943 & {\ul .989} & {\ul .991} & .855 & .919 & {\ul .780} &.786 & .918 & x \\
\textsc{LDM-Re} & \textbf{.943} 	& x	& {\ul .925} & {\ul .944} & \textbf{.990} & \textbf{.992} & {\ul.869} & {\ul .917} & .770 &.787 &.926 & x   \\ \midrule
\textsc{HBDM} & .920  &  .960 &  .917  & {\ul .944}  &  .980 &  .986 &  .853  &  .915 &\textbf{.786} &{\ul .792} & {\ul .919} & {\ul .956} \\
\textsc{HBDM-Re} & .939 & {\ul .964} & \textbf{.926}  & \textbf{.953}  & .986  & .988  & \textbf{.871}  & \textbf{.924} & .774 &\textbf{.795} & \textbf{.930} & \textbf{.963} \\\bottomrule
\end{tabular}
}%
\end{center}
\end{table}

\begin{table}
\caption{AUC for varying representation sizes over the large-scale networks.}
\label{tab:auc_big_networks}
\begin{center}
\resizebox{0.48\textwidth}{!}{%
\begin{tabular}{rcccccccccccc}\toprule
\multicolumn{1}{l}{} & \multicolumn{3}{c}{\textsl{YouTube}} & \multicolumn{3}{c}{\textsl{Flickr}} & \multicolumn{3}{c}{\textsl{Flixster}}& \multicolumn{3}{c}{\textsl{Amazon}}\\\cmidrule(rl){2-4}\cmidrule(rl){5-7}\cmidrule(rl){8-10}\cmidrule(rl){11-13}
\multicolumn{1}{r}{Dimension ($D$)} & $2$ & $3$ & $8$ & $2$ & $3$ & $8$ & $2$ & $3$ & $8$& $2$ & $3$ & $8$ \\\cmidrule(rl){1-1}\cmidrule(rl){2-2}\cmidrule(rl){3-3}\cmidrule(rl){4-4}\cmidrule(rl){5-5}\cmidrule(rl){6-6}\cmidrule(rl){7-7}\cmidrule(rl){8-8}\cmidrule(rl){9-9}\cmidrule(rl){10-10}\cmidrule(rl){11-11}\cmidrule(rl){12-12}\cmidrule(rl){13-13}
\textsc{DeepWalk}    & .822 & .891 & .921 & .889 & .937 & .972 &   .820 &	.866	& .921 &.839 &.932 &.972 \\
\textsc{Node2Vec}    & -     & -     & -     & -     & -     & -     & -     & -     & - &.813 &.880 &.968    \\
\textsc{LINE}        & .660 &  .832  & .878 & .685 & .889 & .812 & .523 & .868 & .936&.626 &.501 &.500 \\
\textsc{NetMF}      & x & x & x & x & x & x & x & x & x &.829 &.831 &.858\\
\textsc{NetSMF}      & .939&  .940 &  .949 & {\ul .974} & .977 & .980 & {\ul.987} & {\ul.987} & {\ul.987}&.768 &.786 &.835 \\
\textsc{RandNE}      & .672 & .700 & .762 & .833 & .869 & .903 & .700 & .739 & .835&.507 &.511 &.514 \\
\textsc{LouvainNE}   & .820 & .819 & .815 & .898 & .899 & .909 & .735 &.718 & .746 &.955 &.954 &.954\\
\textsc{ProNE}  & .691 & .761 & .861 & .623 & .819 & .908 & .756 & .803 & .846   &.847 &.901 &.944  \\
\textsc{Verse}  & \textbf{.957} & \textbf{.964} & \textbf{.971} & .880 & .884 & .858 &\textbf{.988} & \textbf{.988} &\textbf{.988}   &.951 &.977 &{\ul.988}  \\
\midrule
\textsc{HBDM}      &.899     &  .920     & .935      & .972      & {\ul .979}       & {\ul .986}       & .897       & .916       & .932      &{\ul.974} &{\ul.980} &{\ul.988} \\
\textsc{HBDM-Re}   & {\ul.940}      &  {\ul.947}     & {\ul .957}      & \textbf{.980}      & \textbf{.985}       & \textbf{.988}      &   .962    &  .969      &  .971 &\textbf{.976} &\textbf{.981} &\textbf{.989}\\\bottomrule      
\end{tabular}%
}
\end{center}
\end{table}

\begin{figure*}
  \centering
 % \subfloat[(a) \textsl{D=2}]{{ \includegraphics[width=0.28\textwidth]{figures/Micro_dblp_D=2.pdf}\label{fig:dblp_D2} }}
 %  \hfill
 %  \subfloat[(b) \textsl{D=3}]{{ \includegraphics[width=0.28\textwidth]{figures/Micro_dblp_D=3.pdf}\label{fig:dblp_D3} }}\hfill
 %  \subfloat[(c) \textsl{D=8}]{{ \includegraphics[width=0.28\textwidth]{figures/Micro_dblp_D=8.pdf}\label{fig:dblp_D8} }}
 \includegraphics[width=0.9\textwidth]{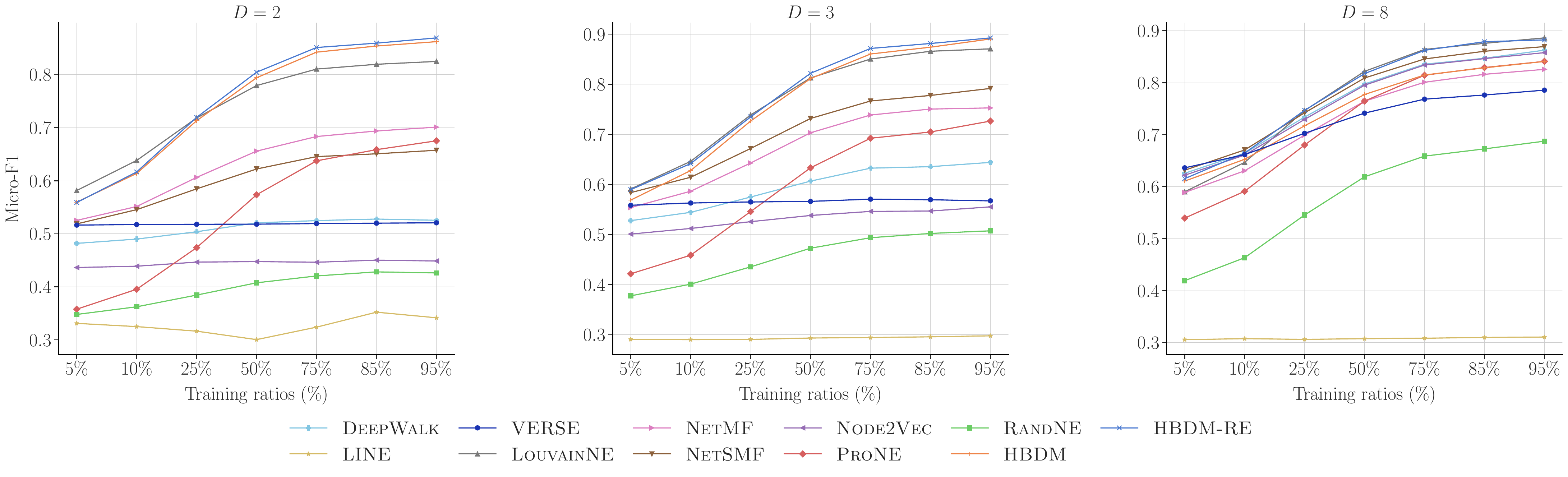}
  \caption{Micro-F1 scores for the classification task considering different low-dimensions and training set ratios for the \textsl{DBLP} network.}\label{fig:microF1_D} 
\end{figure*}

\begin{figure}
    \centering
\includegraphics[width=0.45\textwidth]{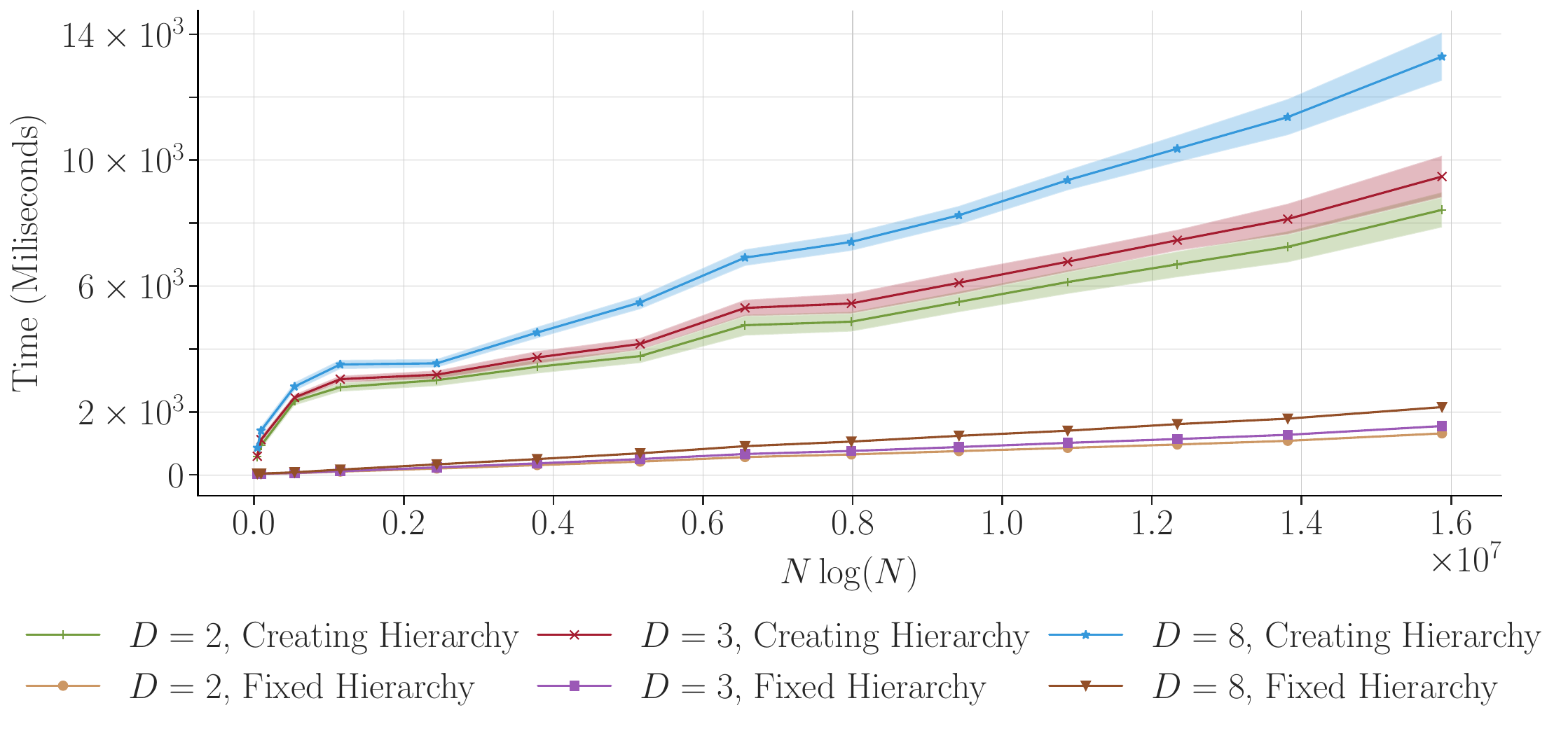}
    \caption{$\textsl{YouTube}$ network---\textsc{HBDM-Re} runtime complexity in miliseconds (ms) as a function of increasing sample sizes of network nodes (in terms of $N\log N$) until the sample set becomes the node set of the total graph. The y-axis showcases the average runtime over $100$ iterations of the forward pass while the shaded areas provide standard deviations, as a measure of uncertainty. Runtimes are presented across $D=2,3,8$ dimensions while we also show the runtime when the inferred hierarchy of the \textsc{HBDM-Re} is created from scratch versus when it is kept static.}
    \label{fig:compl_yes_}
\end{figure}

%\begin{figure}[!b]
%  \centering
  % \subfloat[(a)]{{ \includegraphics[width=0.35\textwidth]{figures/microf1.pdf} }}%
  %\hfill
  %\subfloat[(ii)]{{ \includegraphics[width=0.35\textwidth]{figures/roc_iteration.pdf}\label{fig:roc_iteration} }}%
  %\caption{(i) AUC performance over various networks for varying embedding sizes. 
  %(ii) AUC scores of \textsc{HBDM-Re} in terms of iterations for large-scale networks.}%\label{fig:roc_d} I removed this captions to detect the problems in referencing the figures in text. Some references are wrong because of the split
%\end{figure}

\textbf{Large-Scale Networks:} Results for the large-scale networks are given in TABLE~\ref{tab:auc_big_networks}. Again, we observe that \textsc{HBDM-Re} was on par with the most competitive baselines of \textsc{NetSMF} and \textsc{Verse} while significantly outperforming the rest across networks. We also here find that the inclusion of random effects in the \textsc{LDM}s improves performance highlighting the importance of explicitly accounting for degree heterogeneity also for large networks. 
Notably, we again detect very good performance for the \textsc{HBDM-Re}, but also for \textsc{NetSMF} and \textsc{Verse} when utilizing the very low embedding dimension of $D=2$. %In fact, only very minor improvements are observed for these procedures when increasing the dimensionality to $D=8$ with largest observed improvement found to be only an increase in AUC of $0.017$. 

\begin{figure*}[t]
  \centering
\subfloat[]{{\hspace{0.01\textwidth} }}%
  \hfill
  \subfloat[\textsl{Cora}]{{ \hspace{0.45\textwidth} }}%
  \hfill
  \subfloat[\textsl{DBLP}]{{ \hspace{0.45\textwidth} }}%
  \vfill
  \subfloat[\scriptsize $D$]{{ }}%
  \hfill
  \subfloat[\scriptsize $2$ (TES)]{{ \hspace{0.12\textwidth} }}%
  \hfill
  \subfloat[\scriptsize $128$ (t-SNES)]{{ \hspace{0.12\textwidth} }}%
  \hfill
  \subfloat[\scriptsize $2$ (TES)]{{ \hspace{0.12\textwidth} }}%
   \hfill
  \subfloat[\scriptsize $128$ (t-SNES)]{{ \hspace{0.12\textwidth} }}%
   \vfill
  \subfloat[]{{ \rotatebox{90}{\scriptsize\textsc{Node2Vec}} }}%
  \hfill
  \subfloat[\scriptsize($\text{NR}=0.863$)]{{\includegraphics[width=0.15\textwidth]{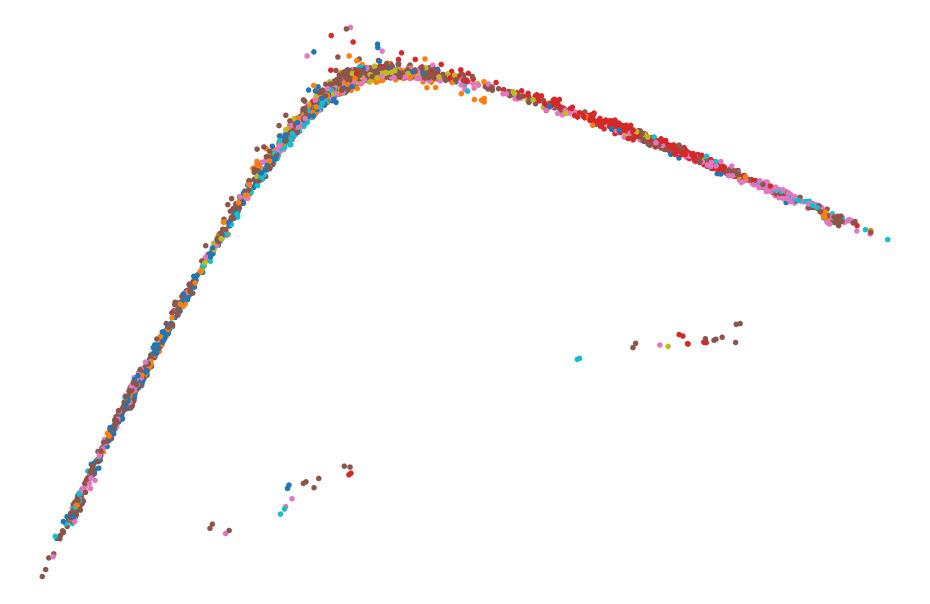} }}%
   \hfill
  \subfloat{{\includegraphics[width=0.15\textwidth]{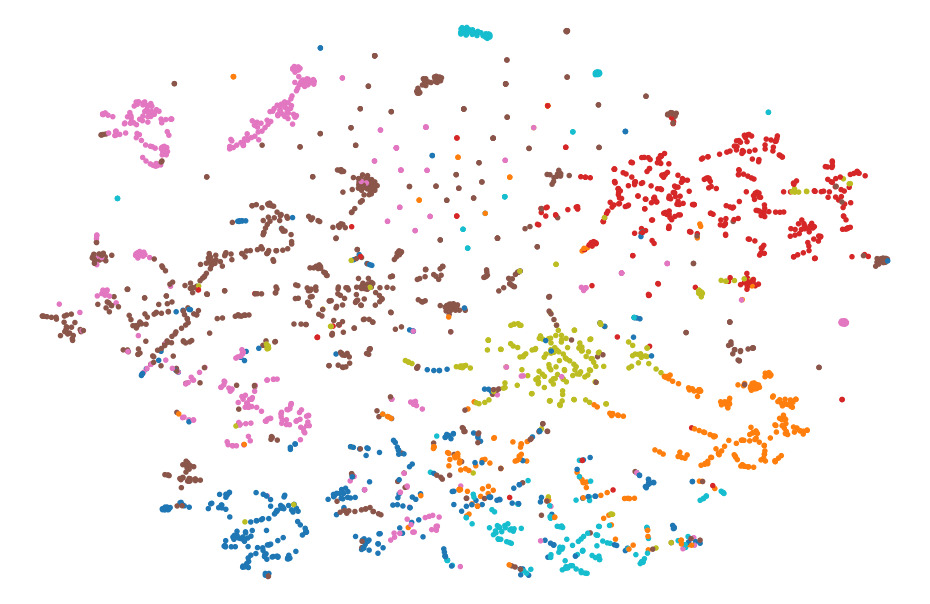} }}%
  \hfill
  \subfloat[\scriptsize($\text{NR}=0.920$)]{{\includegraphics[width=0.15\textwidth]{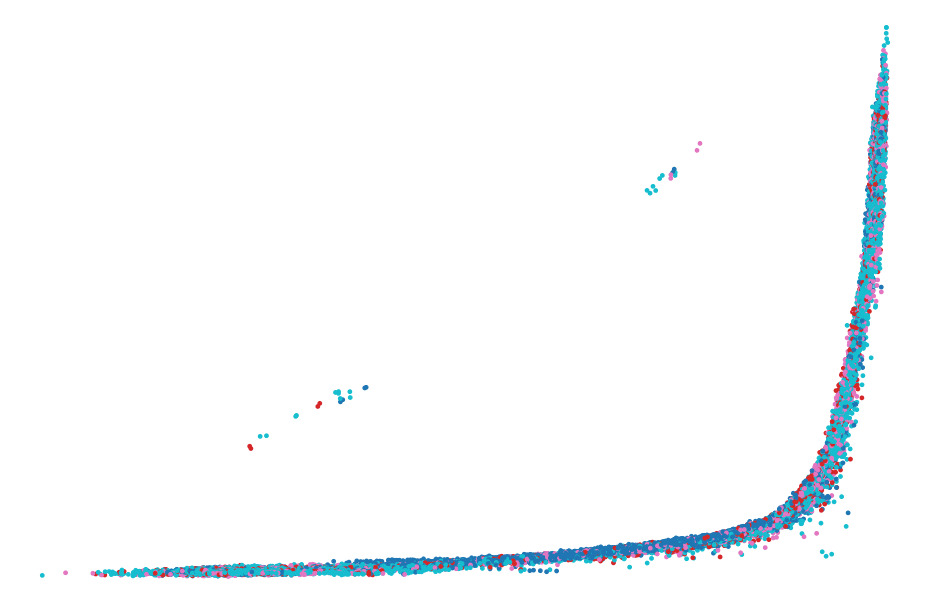} }}%
  \hfill
  \subfloat{{\includegraphics[width=0.15\textwidth]{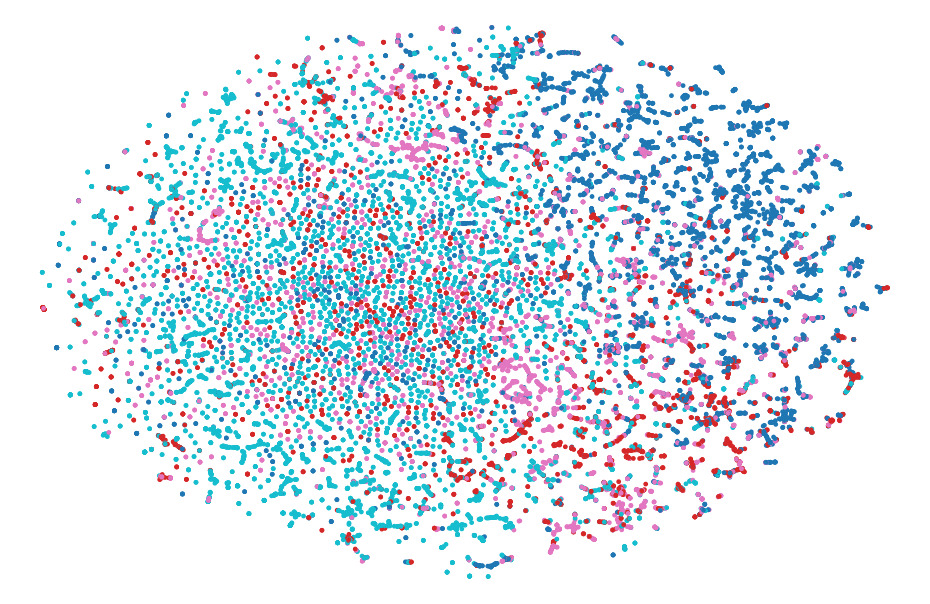} }}%
  \vfill
  \subfloat[]{{ \rotatebox{90}{\scriptsize\textsc{NetSMF}} }}%
  \hfill
  \subfloat[\scriptsize($\text{NR}=0.860$)]{{\includegraphics[width=0.15\textwidth]{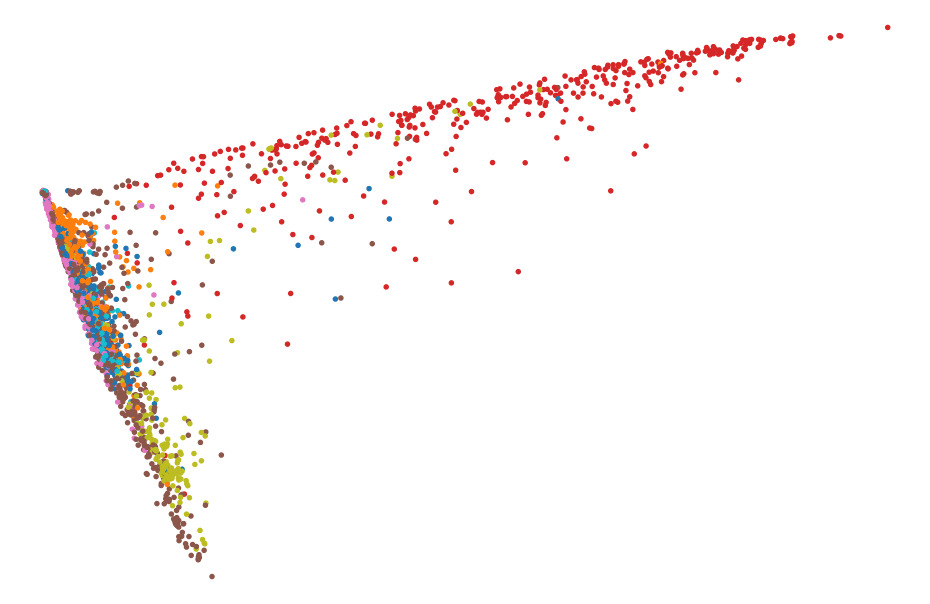} }}%
  \hfill
  \subfloat{{\includegraphics[width=0.15\textwidth]{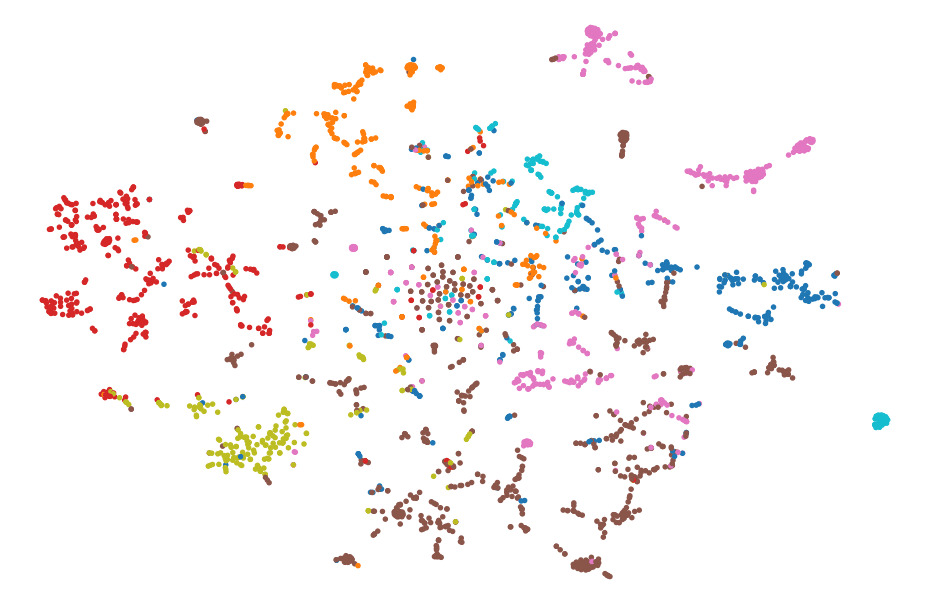} }}%
  \hfill
  \subfloat[\scriptsize($\text{NR}=0.940$)]{{\includegraphics[width=0.15\textwidth]{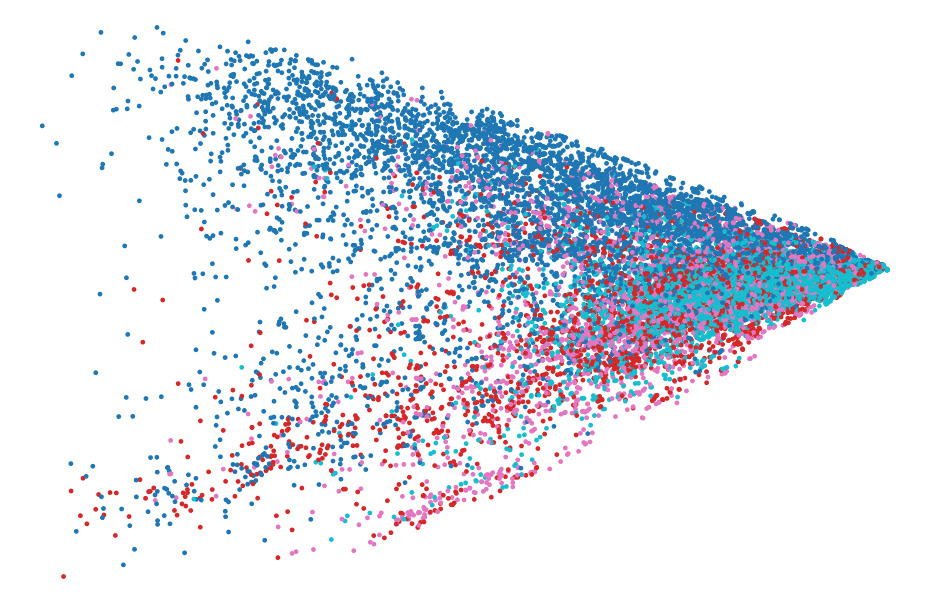} }}%
   \hfill
  \subfloat{{\includegraphics[width=0.15\textwidth]{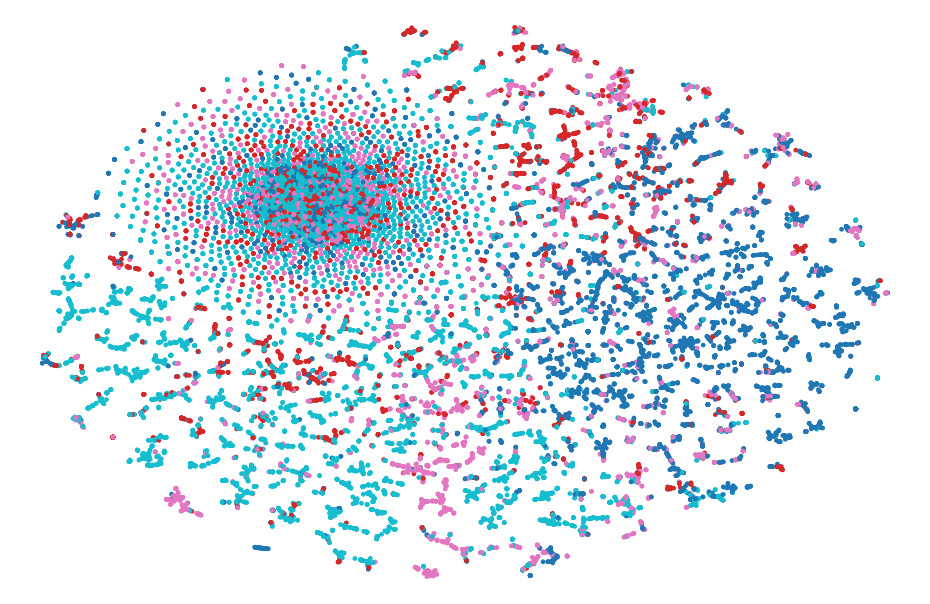} }}%
  \vfill
  \subfloat[]{{ \rotatebox{90}{\scriptsize\textsc{HBDM-Re}} }}%
  \hfill
  \subfloat[\scriptsize($\text{NR}=\mathbf{0.975}$)]{{\includegraphics[width=0.15\textwidth]{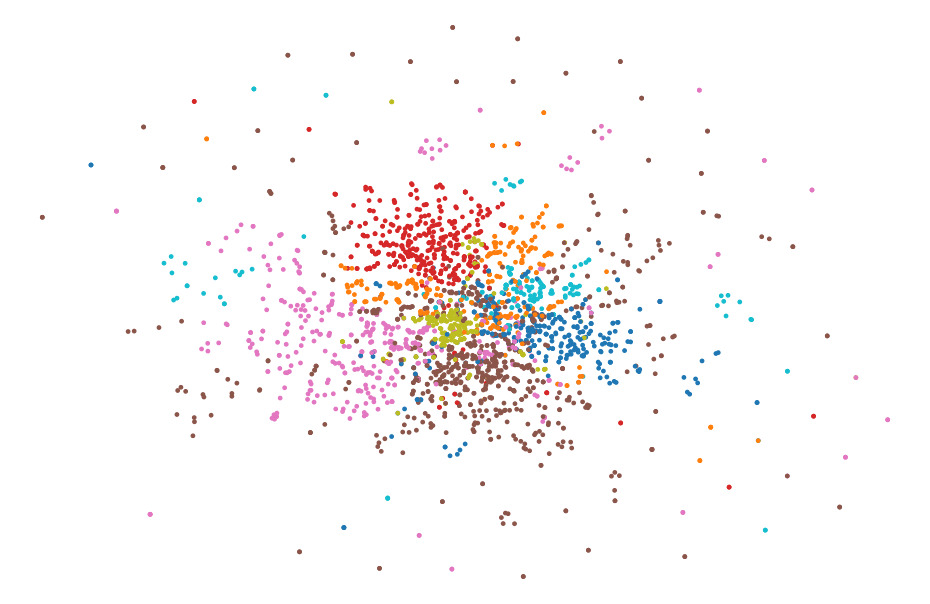} }}%
  \hfill
  \subfloat{{\includegraphics[width=0.15\textwidth]{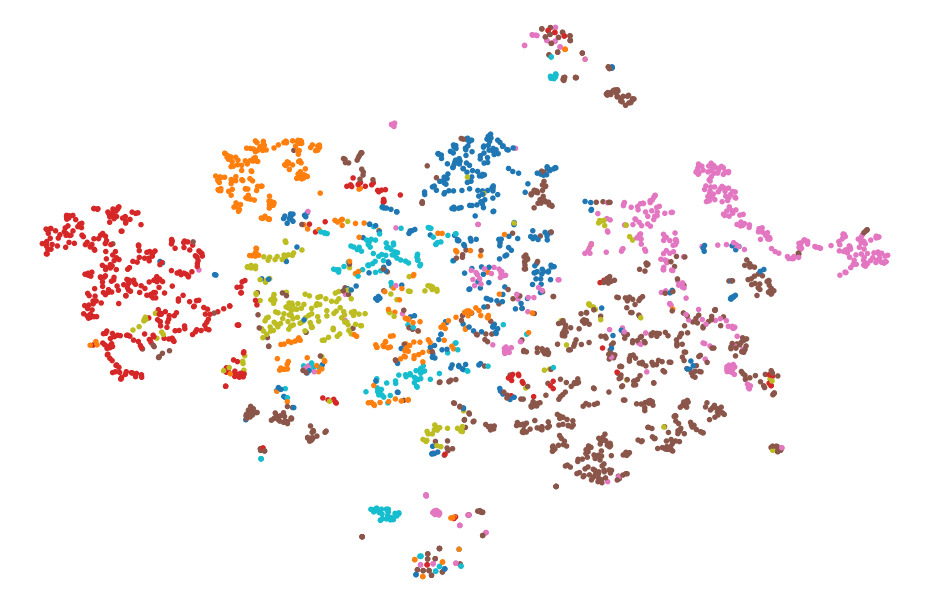} }}%
  \hfill
  \subfloat[\scriptsize($\text{NR}=\mathbf{0.985}$)]{{\includegraphics[width=0.15\textwidth]{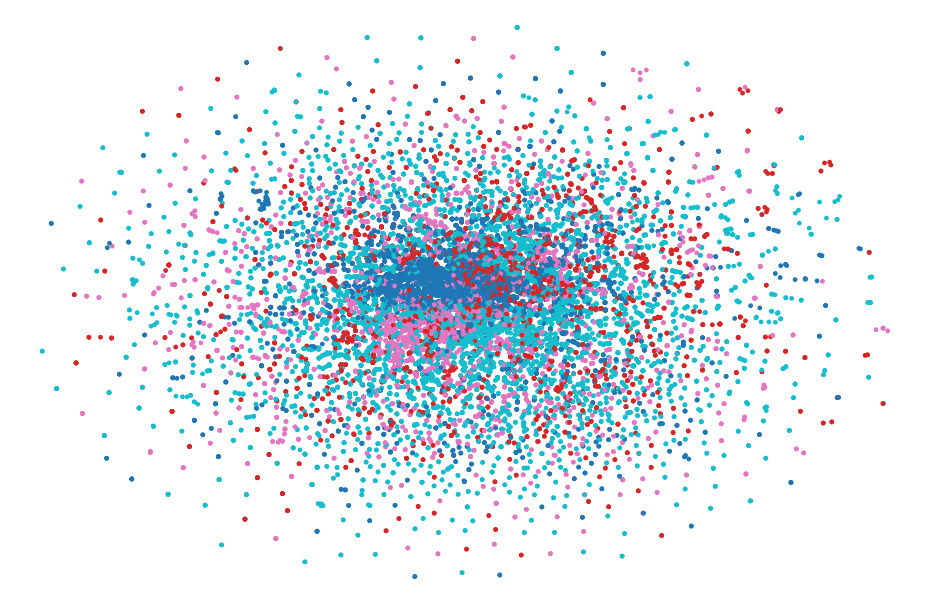} }}%
  \hfill
  \subfloat{{\includegraphics[width=0.15\textwidth]{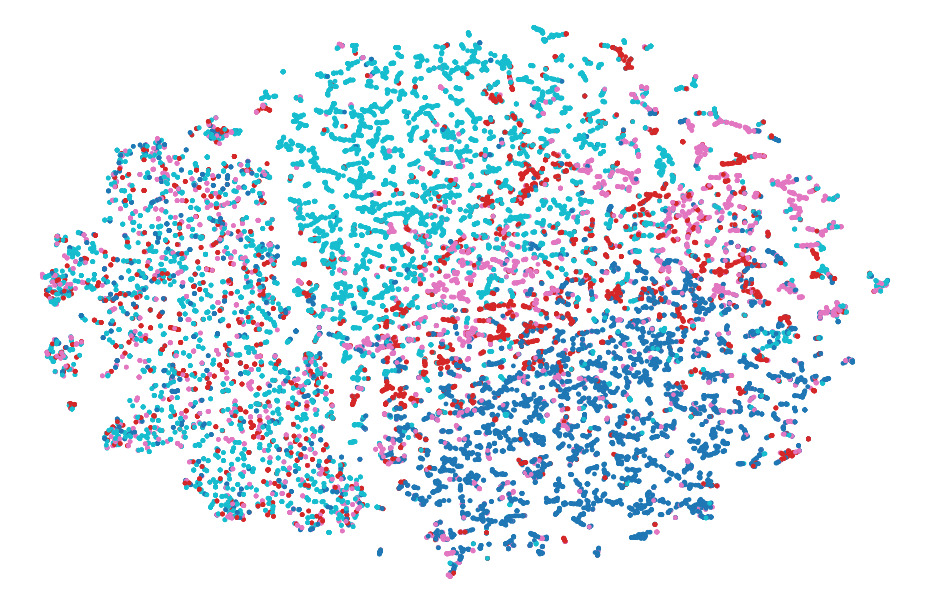} }}%
   \caption{2-D true embedding space versus 128D t-SNE constructed space. For TES, we provide AUC-ROC for the network reconstruction (NR) task.}

  \label{fig:tsne}
\end{figure*}

\textbf{Bipartite Networks:} We validate the performance of our proposed framework for bipartite structures by reporting the AUC score. We perform our experiments on three bipartite networks: (1) \textsl{Drug-Gene} \cite{drug_gene} ($N_1=5,017$, $N_2=2,324$, $|E|=15,138$), (2) \textsl{GitHub} \cite{github} ($N_1=56,519$, $N_2= 120,867$, $|E|=440,237$), and (3) \textsl{Gottron-Reuters} \cite{reuters} ($N_1=21,557$, $N_2=38,677$, $|E|=1,464,182$) following the same experimental setting as in the undirected case of the moderate-sized networks (network details are given in the supplementary). We provide the results in TABLE~\ref{tab:auc-bip} where we witness how the random-effects formulation of \textsc{HBDM-Re} and \textsc{Verse} outperform all the baselines and in most cases significantly. Another interesting observation is that the random-effects model has a notably higher performance than the corresponding global bias model, %This does not come as a surprise 
as the three studied networks have a high degree of heterogeneity.

\subsection{Hyperparemeter sensitivity}
We here study the effect of hyperparameters introduced by the \textsc{HBDM} frameworks. Contrary to many GRL approaches, our models only define three hyperparameters which include the embedding dimensionality $D$, the number of training iterations, and the learning rate $lr$ for the optimizer. In Fig.~\ref{fig:auc_vs_dim}, we view the predictive performance as a function of latent dimension $D$ and here, in general, attain modest improvements in the predictive performance when increasing the embedding dimensions from $D=2$ to $D=8$ with no further improvements increasing to $D=128$, highlighting the efficiency in which \textsc{HBDM} and \textsc{HBDM-Re} utilize very low-dimensional representations. Fig.~\ref{fig:auc_vs_lr}, shows the effect that the learning rate has on performance. We here witness that very small choices $lr\approx 0.001$ define a very slowly increasing performance. Medium magnitude choices of $lr\in [0.005, 0.01]$ define faster convergence while the optimum choices defining very rapid performance saturation exist in the $lr\in [0.05,0.1]$ regime. In Fig.~\ref{fig:roc_iteration}, we investigate the convergence of the best performing \textsc{HBDM-Re} $D=8$ for the large networks, and we witness that the model rapidly converges. After a couple of thousand iterations (very-scalable regime), we already obtain competitive performance for link prediction, which then gently increases until convergence. Our hyperparameter sensitivity analysis focuses on the predictive performance in the downstream task of link prediction. Since our method defines a likelihood over the network, the link predictive performance here shows how well the proposed framework characterizes generalizable patterns of network structure and we therefore focus the analysis on this aspect rather than node classification. If the network structure complies with the node classes, we can expect the node classification performance to follow the same behavior as the link prediction task. %This is a direct consequence of the link prediction task, essentially describing the predictive log-likelihood which expresses the "goodness" of fit to the network. 
Potential disagreement in classification scores against the link prediction scores implies that the node labels do not follow the network structure, and such discrepancy would be network specific rather than a limitation of the method to be investigated.

\begin{table}
\caption{$\text{Micro-F}_1$ scores varying dimension sizes for two moderate and large-scale networks. %The symbol "-" indicates that the running time of the corresponding model takes more than 20 hours and "x" shows that the method is not able to run due to insufficient memory space.
}
\label{tab:micro-f1}
\begin{center}
\resizebox{0.48\textwidth}{!}{%
\begin{tabular}{rcccccccccccc}\toprule
\multicolumn{1}{l}{} & \multicolumn{3}{c}{\textsl{Cora}} & \multicolumn{3}{c}{\textsl{DBLP}} & \multicolumn{3}{c}{\textsl{Amazon}}& \multicolumn{3}{c}{\textsl{YouTube}}\\\cmidrule(rl){2-4}\cmidrule(rl){5-7}\cmidrule(rl){8-10}\cmidrule(rl){11-13}
\multicolumn{1}{r}{Dimension ($D$)} & $2$ & $3$ & $128$ & $2$ & $3$ & $128$ & $2$ & $3$ & $8$& $2$ & $3$ & $8$ \\\cmidrule(rl){1-1}\cmidrule(rl){2-2}\cmidrule(rl){3-3}\cmidrule(rl){4-4}\cmidrule(rl){5-5}\cmidrule(rl){6-6}\cmidrule(rl){7-7}\cmidrule(rl){8-8}\cmidrule(rl){9-9}\cmidrule(rl){10-10}\cmidrule(rl){11-11}\cmidrule(rl){12-12}\cmidrule(rl){13-13}
\textsc{DeepWalk}    & .502	&.712&{\ul .838} & .519	&.605&	.822 &   .231 &	.596	& .929 &.293 &.351 &.413 \\
\textsc{Node2Vec}    & .419&	.658& .835   & .448	&.540	&.815     & .096     & .305 & .895 &- &- &-    \\
\textsc{LINE}        & .197	&.191&.794 & .328 &.294	&.771 & .005 & .003 & .003 &.185 &.134 &.177 \\
\textsc{NetMF}      & .389	&.653& .835 & .654 & .707&.742	 &  x &x   & x  &x &x &x\\
\textsc{NetSMF}      & .554	&.705&\textbf{.842} & .622&.732	&\textbf{.829} &.387  &.649  &.845 & .317&.361  &.397  \\
\textsc{RandNE}      & .271&.337&.731 & .406	&.473	&.718	 & .223 & .411 & .787 &.211 &.226 &.277 \\
\textsc{LouvainNE}   &.804 &{\ul .811} &.801 & .780	&.812	&{\ul.825} & \textbf{.970} &\textbf{.971} & \textbf{.974} &\textbf{.362} &.360 &.359\\
\textsc{ProNE}  & .450	&.611	&.830 & .574	&.634	&.825 & .420 & .750 &  .933   &.218 &.274 &.379  \\
\textsc{Verse}  & .471	&.719	&.828 & .518	&.565	&.757 & .078 & .416& {\ul .949}   &.243 &.305 &.393  \\
\textsc{LDM}  & \textbf{.810}	&.802	&.774 & x	&x	&x & x & x& x   &x &x &x  \\
\textsc{LDM-Re}  & .802	&.803	&.796 & x	&x	&x & x & x& x   &x &x &x  \\\midrule
\textsc{HBDM}      &.789 & .807 & .816 &{\ul .812} & {\ul.814} & .772 & \textbf{.970}       & \textbf{.971}      & .931      &.320 &{\ul .366} &{\ul.414} \\
\textsc{HBDM-Re}   & {\ul .805}&\textbf{.813}&  .818  & \textbf{.805}	&\textbf{.822}	& .808 &  {\ul .956} & {\ul .955}  & .931 &{\ul.326} &\textbf{.367} &\textbf{.414}\\\bottomrule    
\end{tabular}%
}
\end{center}
\end{table}

\subsection{Node classification}
We assess the performance of the proposed framework in the uni/multi-label classification task and provide the $\text{Micro-F}_1$ scores in TABLE~\ref{tab:micro-f1} ($\text{Macro-F}_1$ scores are reported in the supplementary). Scores are defined as the mean value over $10$ random shuffles defining the training and test sets. Standard deviations as error bars were found in the scale of $10^{-3}$ and thus not presented. For the experimental setup, we randomly pick $50\%$ of nodes as the training set and use the rest as the testing set. For an accurate comparison across different methods, we used two simple classifiers, a linear (logistic/multinomial regression classifier) and a non-linear (linearithmic k-nearest neighbors (\textsl{kNN}) classifier), and report the highest scores. We found that all methods benefit from using \textsl{kNN}. The number of neighbors was set to $k=10$ (similar results were obtained with higher choices for $k$ as well). Lastly, we report the average $\text{Micro-F}_1$ scores across $10$ repeated trials. Results for the uni-labeled \textsl{Cora} and \textsl{DBLP} networks are reported in the two leftmost columns of TABLE \ref{tab:micro-f1}. We observe that \textsc{HBDM-Re} and \textsc{HBDM} significantly outperform the baselines for the regimes of $D=2,3$ with only $\textsc{LouvainNE}$ being competitive. Results for large-scale and multi-labeled networks \textsl{Amazon} and \textsl{YouTube} are provided by the two rightmost columns in TABLE \ref{tab:micro-f1}. Again, the proposed framework outperforms the baselines for the low-dimensional regime with $\textsc{LouvainNE}$ being on par. Comparing our framework with the classic \textsc{LDM-Re} and \textsc{LDM}, we again see an on-par performance which we attribute to the \textsc{HBDM} well preserving the intrinsic properties of homophily and transitivity. We further investigate the effect that the amount of training data has on classification performance. In Fig. \ref{fig:microF1_D} we provide the performance across multiple training size ratios and consider ultra-low dimensional embeddings of $D=2,3,8$ for the \textsl{DBLP} network. We here observe that for the cases of $D=2,3$, our frameworks significantly outperform all the baselines with only \textsc{LouvainNE} being competitive. Increasing the dimensionality to $D=8$, the baseline models are defined with enough capacity to be competitive while \textsc{HBDM} and \textsc{HBDM-Re} return favorable results.

\begin{table}
\caption{AUC scores for varying representation sizes over three bipartite networks.}
\label{tab:auc-bip}
\begin{center}
\resizebox{0.48\textwidth}{!}{%
\begin{tabular}{rccccccccc}\toprule
\multicolumn{1}{l}{} & \multicolumn{3}{c}{\textsl{Drug-Gene}} & \multicolumn{3}{c}{\textsl{GitHub}} & \multicolumn{3}{c}{\textsl{Gottron-Reuters}}\\\cmidrule(rl){2-4}\cmidrule(rl){5-7}\cmidrule(rl){8-10}
\multicolumn{1}{r}{Dimension ($D$)} & $2$ & $3$ & $8$ & $2$ & $3$ & $8$ & $2$ & $3$ & $8$ \\\cmidrule(rl){1-1}\cmidrule(rl){2-2}\cmidrule(rl){3-3}\cmidrule(rl){4-4}\cmidrule(rl){5-5}\cmidrule(rl){6-6}\cmidrule(rl){7-7}\cmidrule(rl){8-8}\cmidrule(rl){9-9}\cmidrule(rl){10-10}
\textsc{DeepWalk}    & .673	&.843 &.878 & .762	&.853&	.902 &   .673 &	.769	& .905  \\
\textsc{Node2Vec}    & .758 &	.814& .793    & .724	&.823	&.876     & .694     & .766 & .830    \\
\textsc{LINE}        & .798	&.836&.867 & .805 &.766	&.902 & .715 & .696 & .850  \\
\textsc{NetMF}      & .576	&.598 & .742 & .711 & .711 &.708	 &  .747 &.747   & .730  \\
\textsc{NetSMF}      &  .839	&.838  &.796 & .846  &.847	&.857 &.874  &.934  & .941  \\
\textsc{RandNE}      & .536 & .551 &.613 & .615	&.651	&.707	 & .769 & .808 & .872  \\
\textsc{LouvainNE}   & .760 &.767 &.779 & .694	&.702	&.735 & .654 & .648 & .679 \\
\textsc{ProNE}  & .667	&.765	&.831 & .676	&.771	&.840 & .606 & .725 & .909    \\
\textsc{Verse}  & \textbf{.910}	& \textbf{.913}	& \textbf{.922} &  \textbf{.943}	& \textbf{.952}	& \textbf{.959} &  {\ul.962} & {\ul .966} & {\ul .967} \\
\midrule
\textsc{HBDM}      &.798 &.836 & .889 &.849 &  .869&  .905 &  .941       &  .949      & .950      \\
\textsc{HBDM-Re}   & {\ul.872} &{\ul.891}&  {\ul.914}  & {\ul.932}	&{\ul.934}	& {\ul.937} &  \textbf{.964} & \textbf{.967}  & \textbf{.973} \\
%\textsc{HBDM-Re-sq}   & & &    & 	&	&  &   &   & \\
\bottomrule    
\end{tabular}%
}
\end{center}
\end{table}

\begin{figure*}[t]
  \centering
  \subfloat[(i) Dendrogram]{{ \includegraphics[scale=0.08]{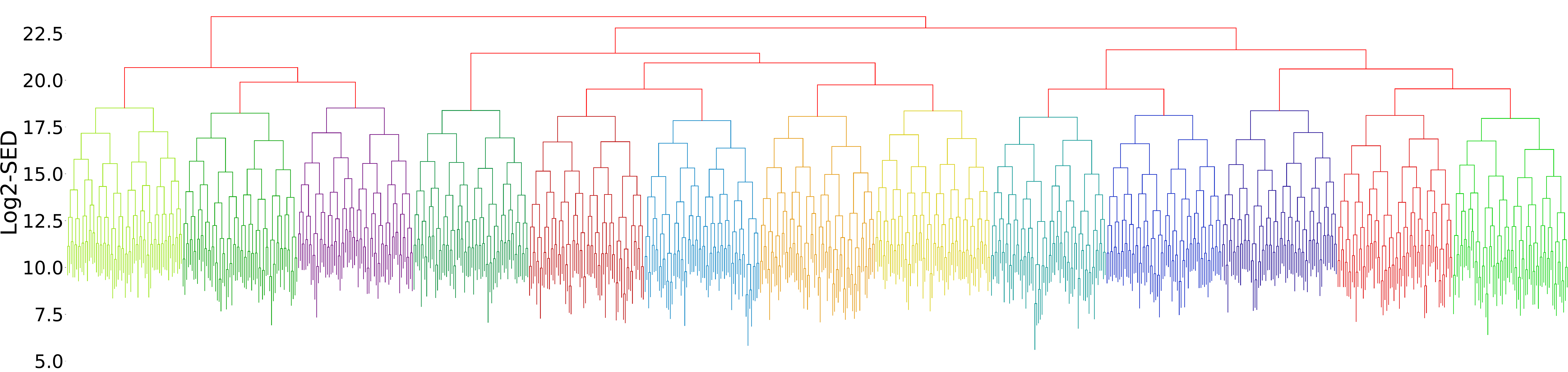} }}%
  \hfill
   \subfloat[(ii) Embedding Space]{{ \includegraphics[scale=0.2]{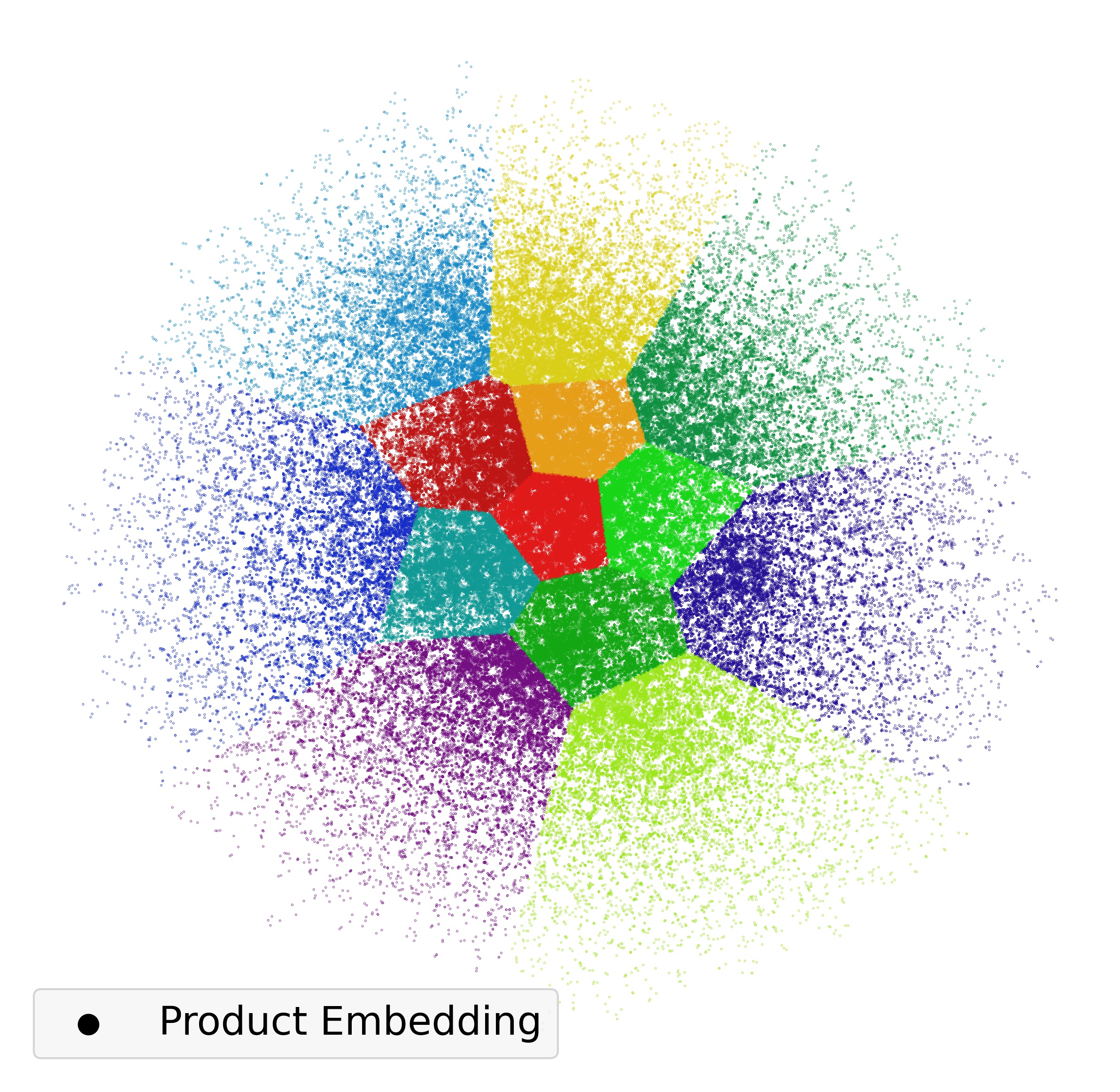} }}%
  \hfill
  \subfloat[(iii) L=1]{{ \includegraphics[scale=0.18]{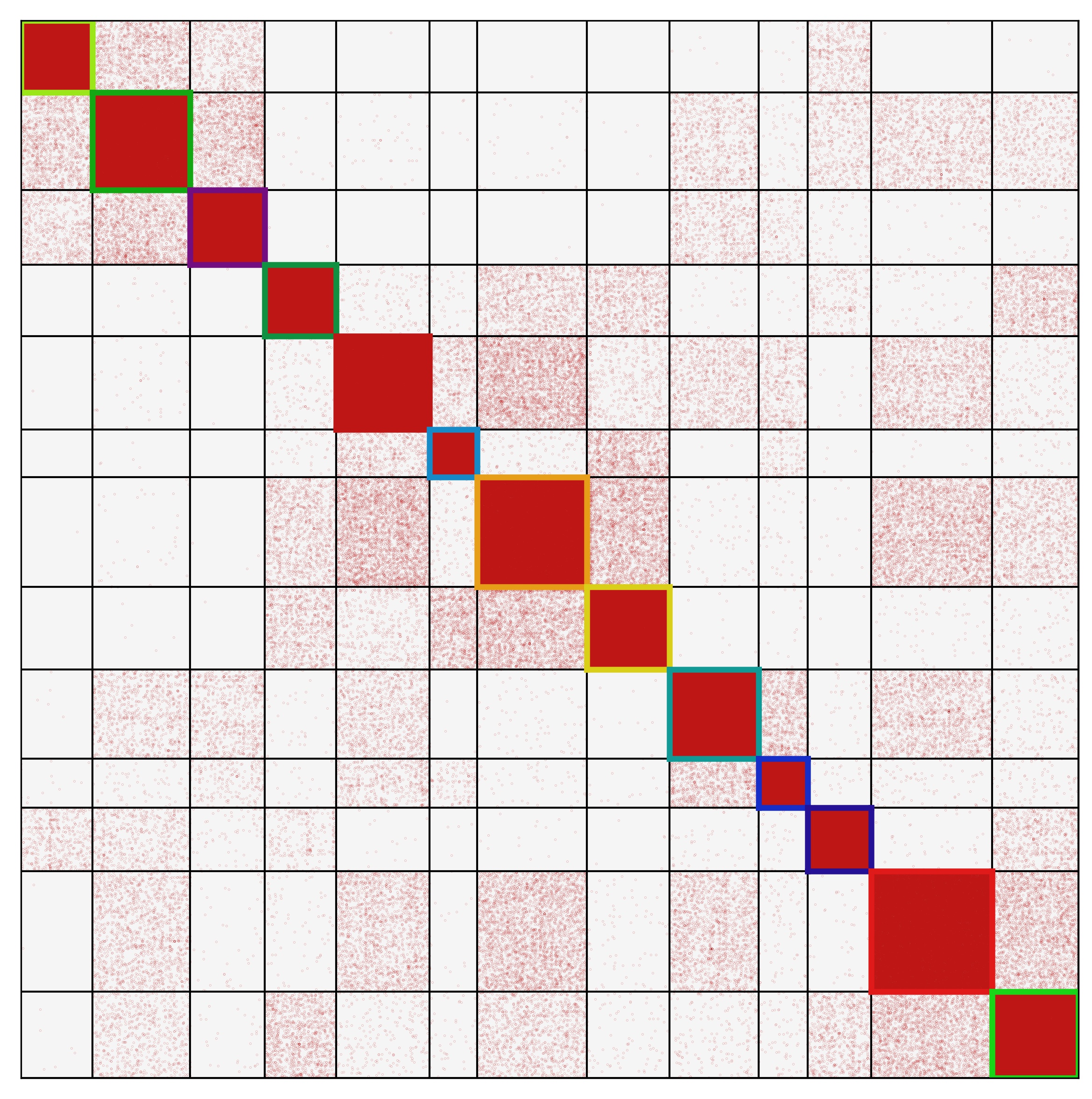} }}%
  \hfill
  \subfloat[(iv) L=3]{{ \includegraphics[scale=0.18]{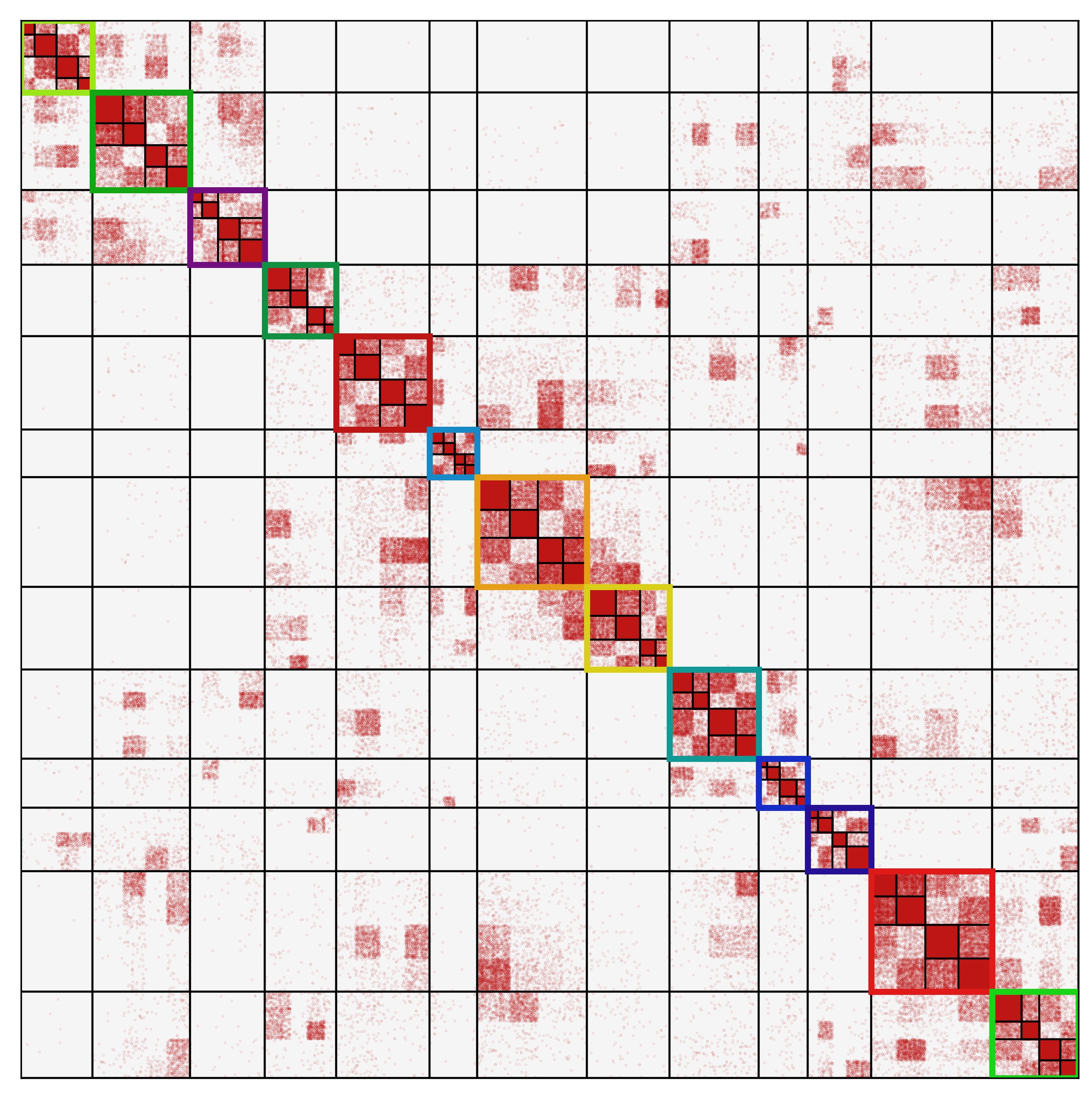} }}%
  \hfill
    \subfloat[(v) L=5]{{ \includegraphics[scale=0.18]{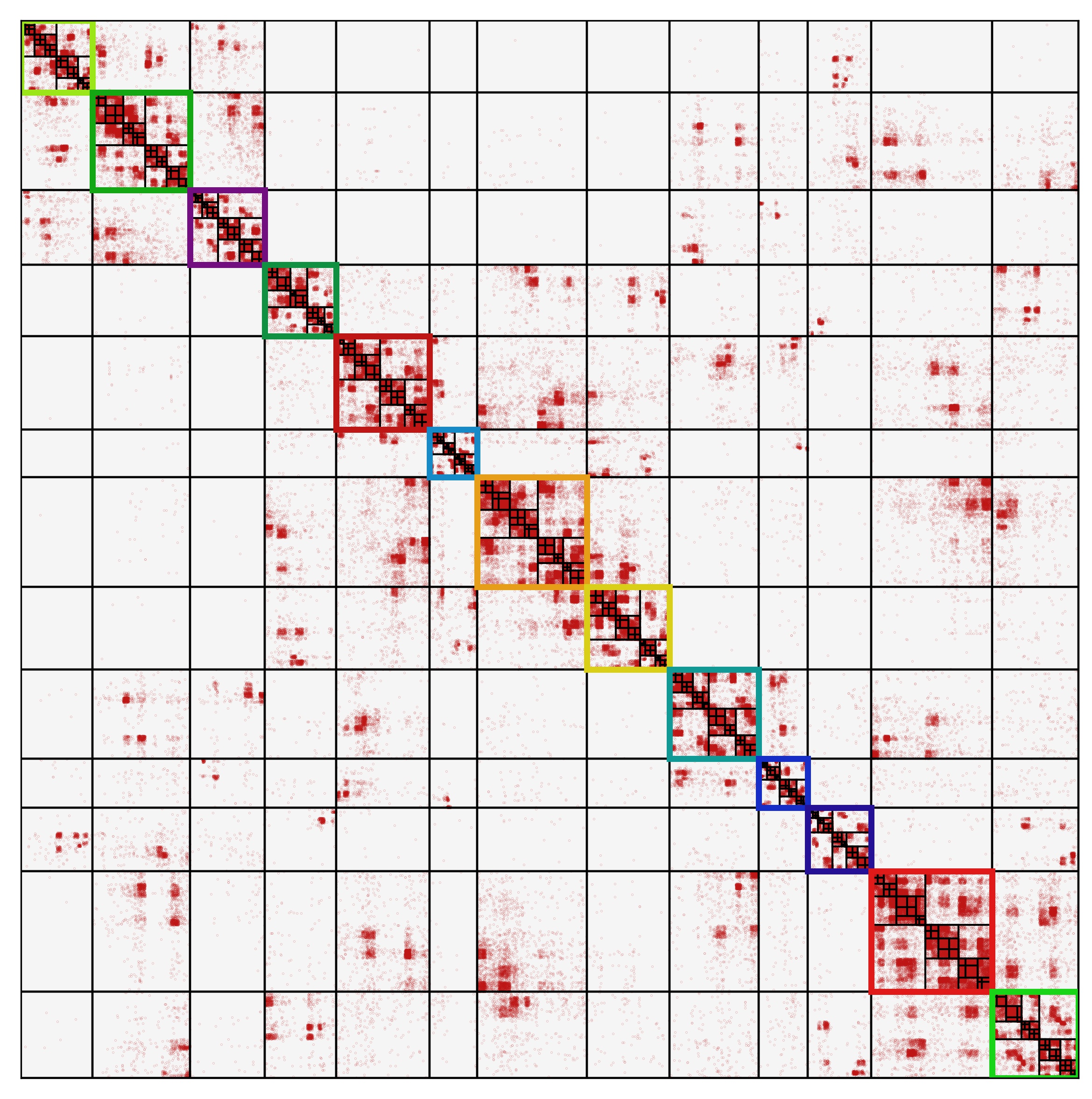} }}%
  \hfill
  \subfloat[(vi) L=8]{{ \includegraphics[scale=0.18]{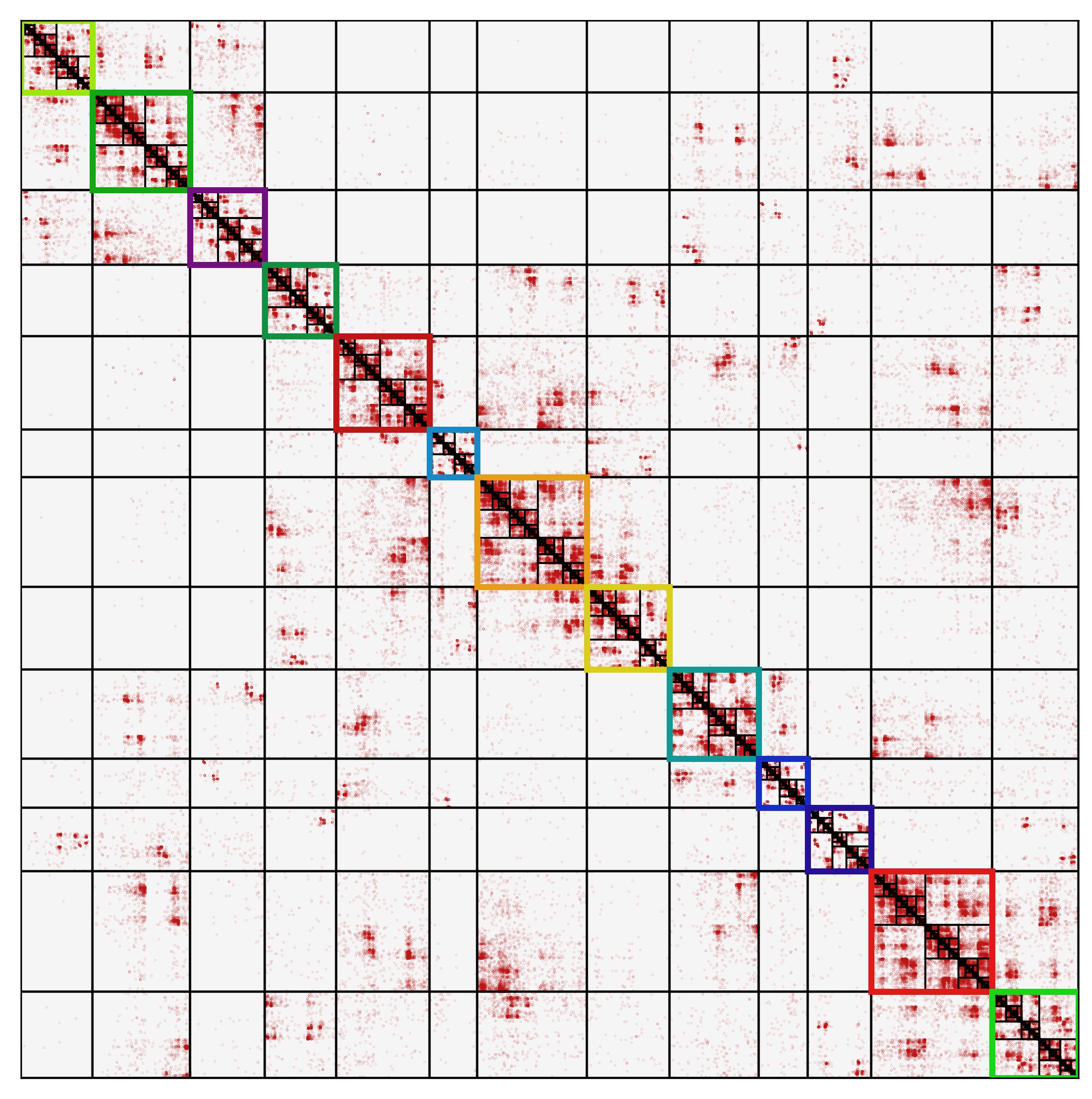} }}%
   \caption{\textsl{Amazon} network dendrogram, embedding space and ordered adjacency matrices for the learned $D=2$ embeddings of \textsc{HBDM-Re} and various levels $(L)$ of the hierarchy. }\label{fig:amazon_dend}
\end{figure*}

\subsection{Across tasks comparison:} We have considered multiple downstream tasks in each of which various baselines were found to be competitive against our \textsc{HBDM} frameworks. In general, the \textsc{HBDM} is characterized by the most consistent performance across tasks, especially for low dimensions. \textsc{NetSMF} was most competitive in the large-scale networks but underperformed in the moderate-sized networks and node classification. \textsc{Verse} was the most competitive baseline across tasks but massively underperformed in node classification for low dimensions. Furthermore, \textsc{LouvainNE} had high performance in node classification but underperformed in link prediction. Models such as \textsc{NetSMF} and \textsc{Verse} can express structural (stochastic) equivalence while our \textsc{HBDM} explicitly expresses homophily and transitivity. This can explain the occasional higher performance of these baselines in the link prediction task.

\subsection{Runtime complexity:}
We assess the runtime complexity of the \textsc{HBDM-Re} framework in terms of increasing network sizes. In Fig. \ref{fig:compl_yes_}, we consider the \textsl{YouTube} network and show the runtime complexity in milliseconds (ms) as a function of increasing sample sizes of network nodes (in terms of $N\log N$) until the whole network is recovered. Runtimes are presented as the average across $100$ iterations of the forward pass while the shaded areas provide the standard deviation. Runtimes are presented across $D=2,3,8$ dimensions while we also compare the runtimes when the \textsc{HBDM-Re} builds the
hierarchical structure via the k-means procedure and when the hierarchical structure is kept fixed. We here observe, that the runtime increases almost linearly as we increase the number of nertwork nodes. Comparing the runtime when creating the hierarchy via the k-means procedure from scratch versus keeping the dendrogram static from past iterations shows a significant decrease in runtime for the latter (in the experiments we create the hierarchy every $25$'th iteration). Thus, the main bottleneck of the \textsc{HBDM-Re} approach is the computations required for the proposed Euclidean k-means procedure. Despite being deemed outside of the scope of this paper, such a bottleneck can be addressed by exploring existing procedures scaling conventional squared Euclidean k-means by avoiding unnecessary distance calculations \cite{elkan2003using} or by the use of binary space partitioning trees \cite{pettinger2010space}. Such improved scaling would even admit utilization of non-binary splits beyond the root node improving the accuracy of the hierarchical approximation.

\subsection{Network visualization} 
The graph representation learning literature mainly focuses on embeddings with dimensionality greater than $D=2 \text{ and } 3$. As a direct consequence, network visualizations rely on dimensionality reduction frameworks, typically using the t-distributed Stochastic Neighbor Embedding (t-SNE) \cite{tsne-laurens08}. In order to verify the quality of the t-SNE constructed Space (t-SNES), we provide the labeled-colored True Embedding Space (TES) in Fig. \ref{fig:tsne} for $D=2$, as well as for $D=128$ mapped to $D=2$ via the use of t-SNE for \textsl{Cora} and \textsl{DBLP}. We see that the \textsc{HBDM-Re} frameworks provide highly informative embeddings with no need for dimensionality reduction, unlike the rest of the baselines. This is also verified from the optimal performance in network reconstruction, \textsc{HBDM-Re} can successfully express the network structure using just $D=2$. %We attribute the increase in class separability while using high-dimensions and t-SNE (especially for \textsl{DBLP}) to the fact that peripheral nodes acting as noise are grouped in a unique cluster, as it is evident from the sixth column in Figure \ref{fig:tsne}. 
%We note here \textsc{HBDM} has similar performance as \textsc{HBDM-Re} and thus left for the supplementary. The visualization experiments position our proposed models as the silver lining between consistent performance across all GRL downstream tasks as well as network visualizations in very low dimensions. All baseline results are available in the supplementary. 
In Fig. \ref{fig:amazon_dend} we provide the hierarchical block structure constructed by the \textsc{HBDM-Re} for the \textsl{Amazon} network. For visualization, we used the average within-cluster Euclidean distance to the centroid $\Big(\Delta\{A,B\}=\frac{1}{N_A+N_B}\sum_{i\in C_A,C_B}\|\bm{z}_i-\bm{\mu}_{A\bigcup B}\|_2\Big)$, as a linkage function to form a post-processing agglomerative clustering, for ordering the initial $\log N$ centroids. In Fig. \ref{fig:amazon_dend} (i), we provide the dendrogram which denotes the agglomeration result in the top-level with red lines. The dendrogram continues with the hierarchical splits of our \textsc{HBDM-Re} where each color indicates the initial $\log N$ blocks. The y-axis of the dendrogram represents the binary logarithm of the Sum of Euclidean Distances, $\textsl{Log2-SED}=\log_2\Big(\sum_{i\in C_k^{(l)}}\|\bm{z}_i-\bm{\mu}_{k}^{(l)}\|_2\Big)$. Moreover, Fig. \ref{fig:amazon_dend} (ii) conveys the corresponding latent space, colored based on the coarse $\log N$ split, revealing directly interpretable and accurate network representations. In Fig. \ref{fig:amazon_dend} (iii), (iv), (v) and (vi) we showcase the organized adjacency matrices, based on the 2-dimensional \textsc{HBDM-Re} learned hierarchy for various levels $L$ of the tree. We here, observe %the increasing proximity between centroids as we move down the hierarchy, as well as, 
the representation power of the extracted hierarchy from just a $2$-dimensional \textsc{HBDM-Re} defining communities and their sub-communities at finer and finer details. 

\begin{figure*}[]
  \centering
  \subfloat[(i) Dendrogram]{{ \includegraphics[scale=0.08]{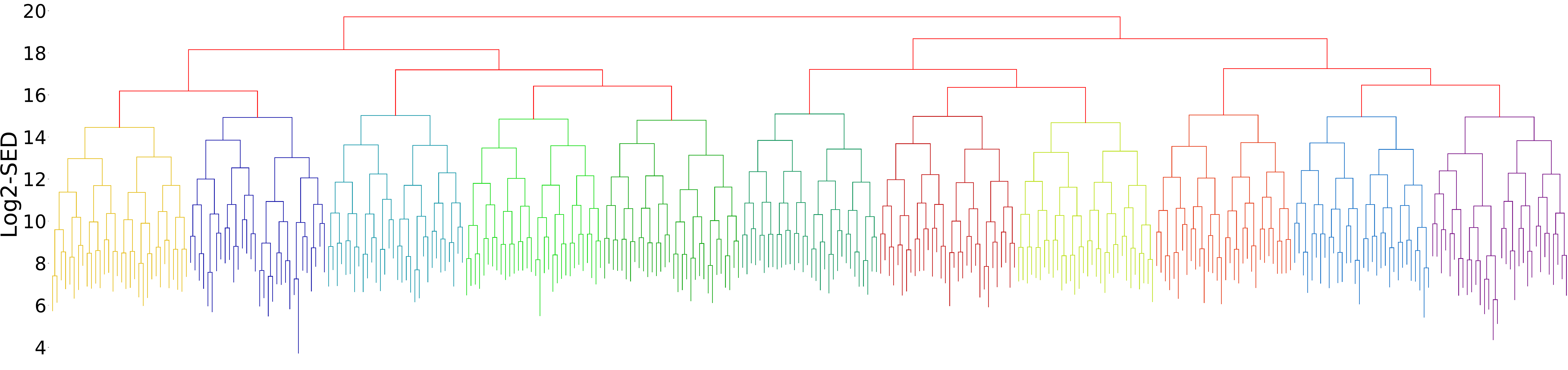} }}%
  \hfill
  \subfloat[(ii) Embedding Space]{{ \includegraphics[scale=0.2]{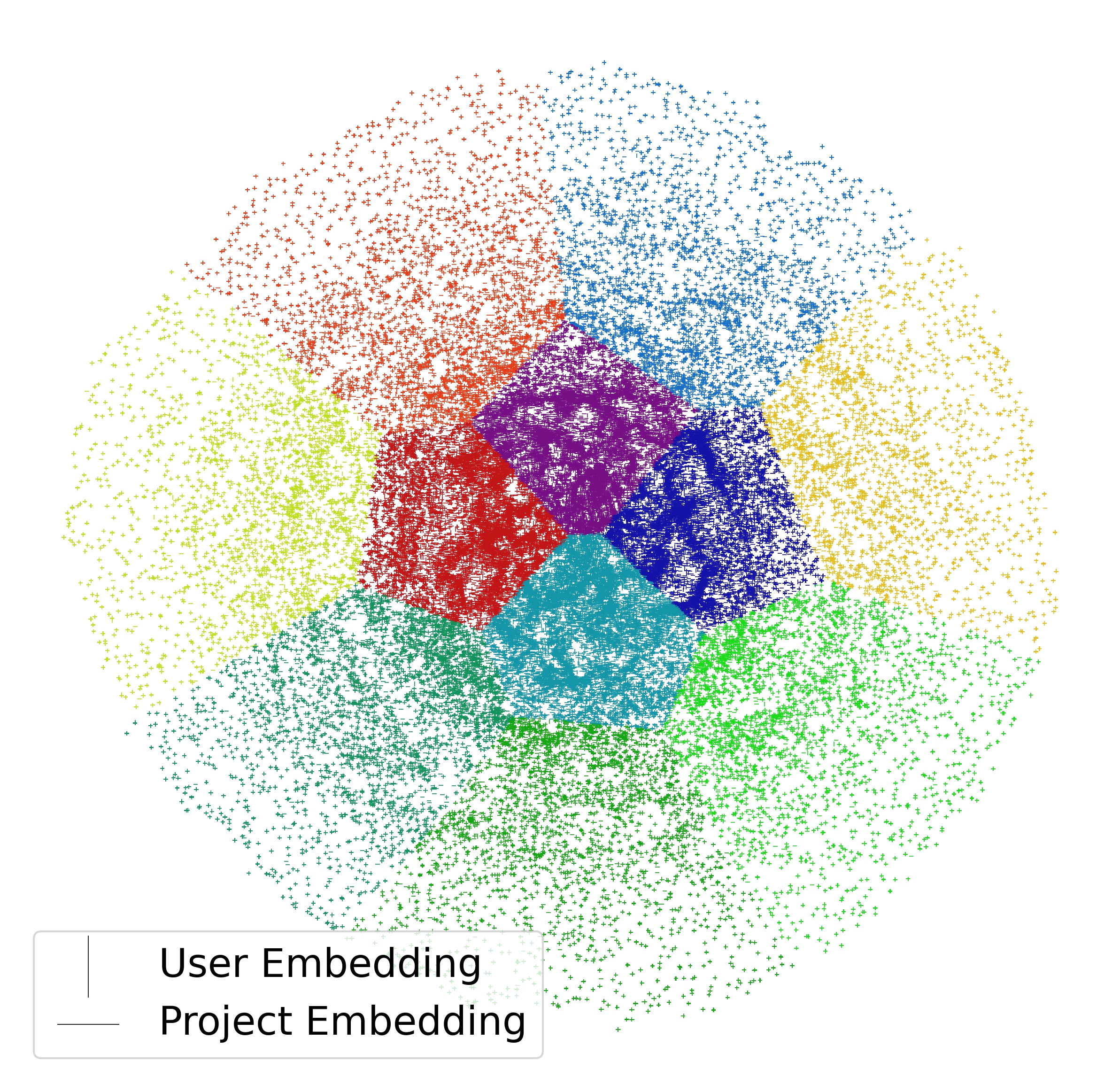} }}%
  \hfill
  \subfloat[(iii) L=1]{{ \includegraphics[scale=0.27]{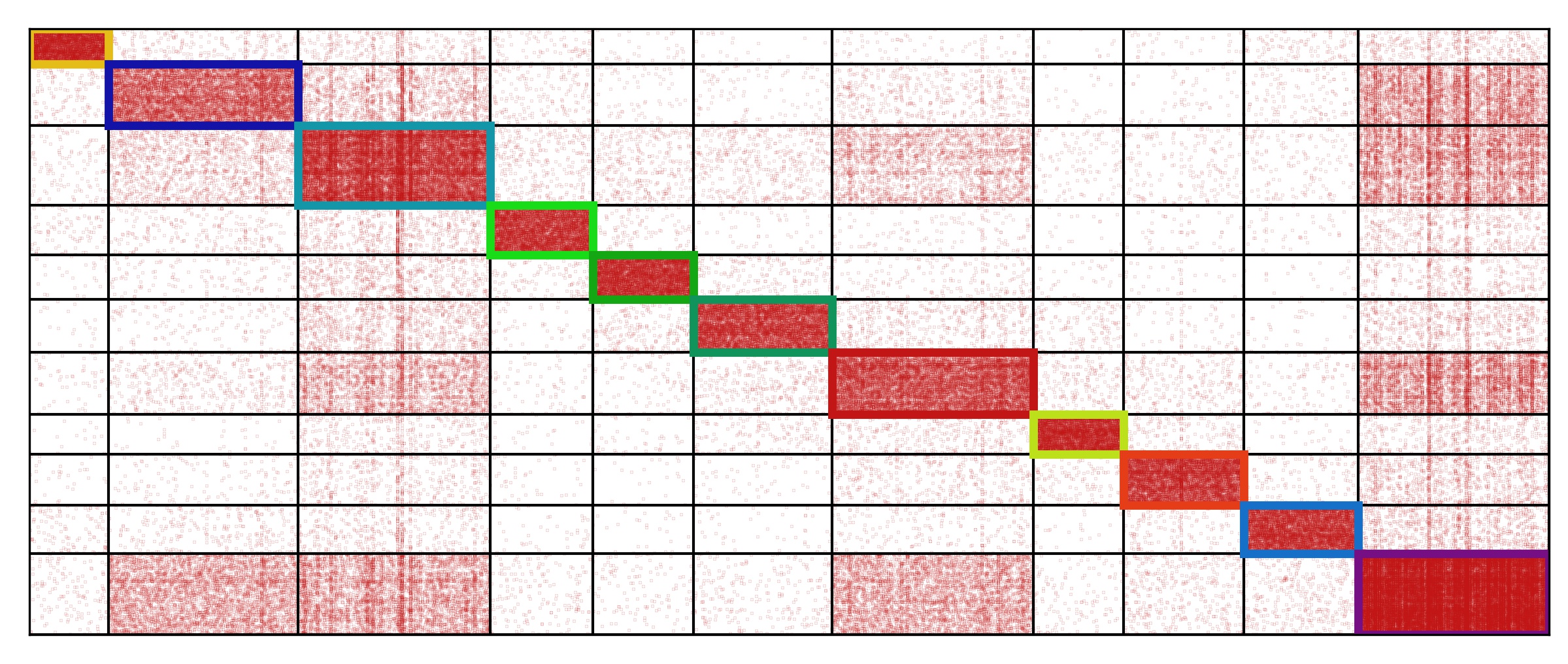} }}%
  \hfill
  \subfloat[(iv) L=3]{{ \includegraphics[scale=0.27]{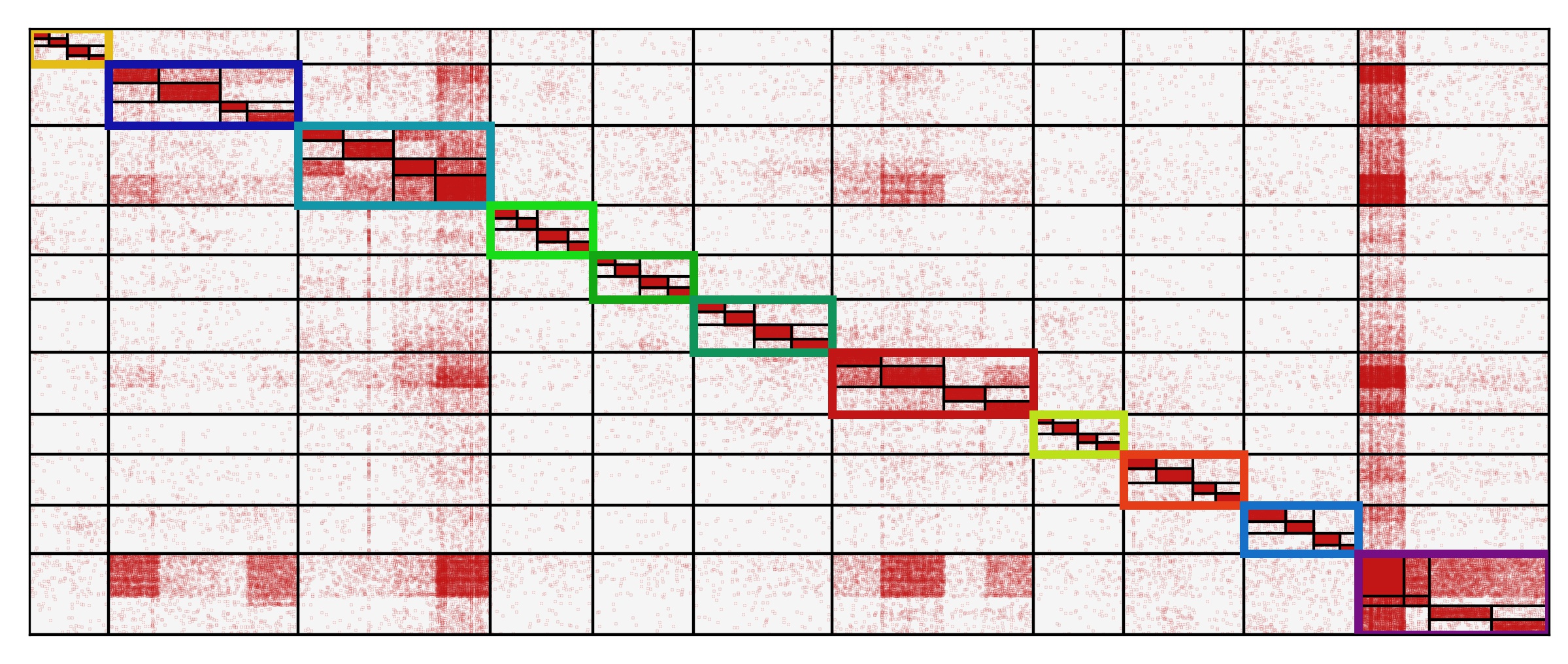} }}%
  \hfill
  \subfloat[(v) L=6]{{ \includegraphics[scale=0.27]{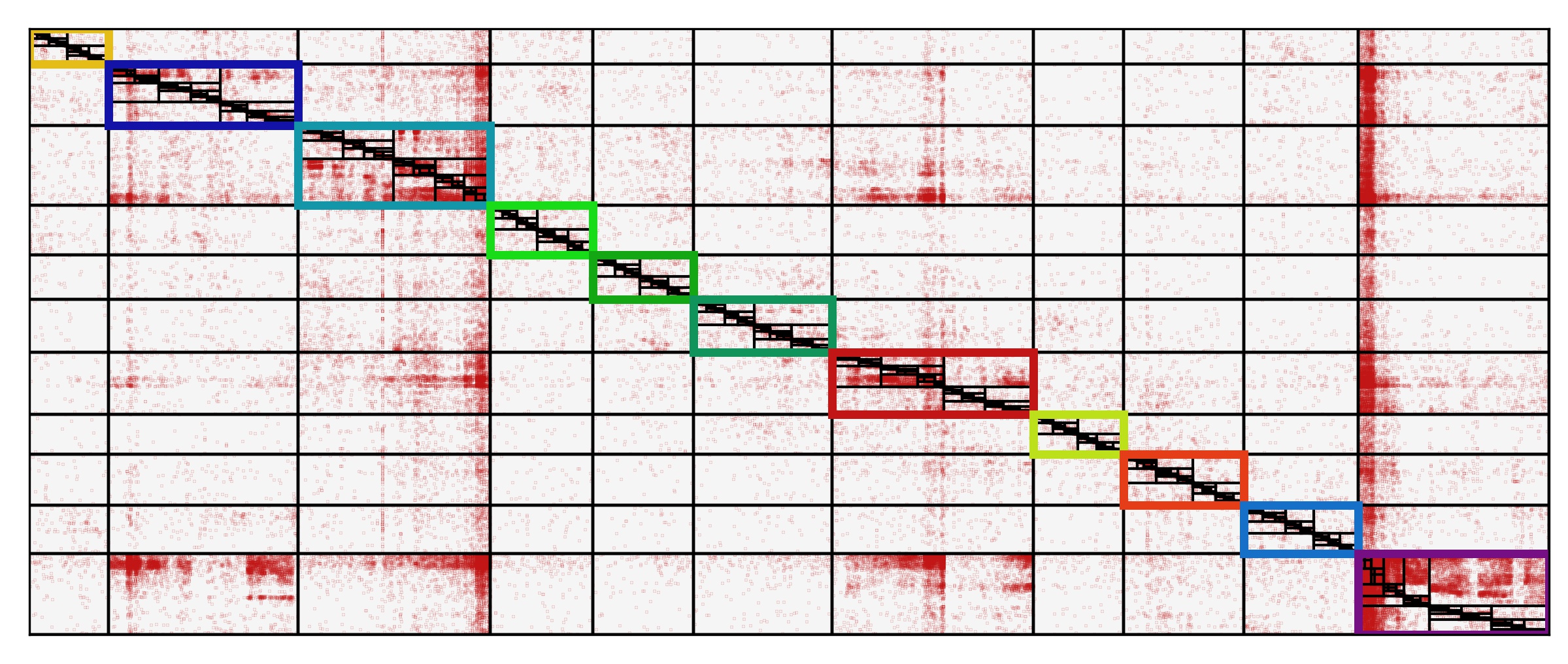} }}%
  \caption{\textsl{GitHub} network dendrogram, embedding space and ordered adjacency matrices for the learned $D=2$ embeddings of \textsc{HBDM-Re} and various levels $(L)$ of the hierarchy. }\label{fig:github_dend}
\end{figure*}

For the bipartite case, we show how \textsc{HBDM-Re} can enhance our understanding of the bipartite structure at multiple scales and levels. Similar to the undirected case, Fig. \ref{fig:github_dend} (i), indicates the dendrogram of the imposed hierarchy, enriched with agglomeration for a coarse level block ordering and proximity for the \textsl{GitHub} network. In addition, Fig. \ref{fig:github_dend} (ii), provides the corresponding latent space, colored based on the coarse $\log N$ split. Notably, no dimensionality-reduction is necessary to define accurate network representation in the latent space of the two disjoint populations and visually access and express node similarity. In Fig. \ref{fig:github_dend} (iii), (iv) and (v), we exhibit %for the \textsl{GitHub} network (additional networks are provided in the supplementary)
how the multi-scale structure evolves through different levels of the hierarchy defined by \textsc{HBDM-Re}, showcasing how a joint bi-clustering for complex network embeddings naturally can be obtained, with no need for post-processing steps. Our \textsc{HBDM}, can thus accurately characterize bipartite networks and successfully uncover their hierarchical block structure efficiently.

%\begin{figure}
 % \centering
 % \subfloat[(i) Dendrogram]{{ \includegraphics[scale=0.08]{figures/bipartite_networks/drug_gene.pdf} }}%
 % \hfill
 % \subfloat[(ii) L=1]{{ \includegraphics[scale=0.08]{figures/bipartite_networks/drug_gene_l=0.jpeg} }}%
 % \hfill
  %\subfloat[(iii) L=2]{{ \includegraphics[scale=0.08]{figures/bipartite_networks/drug_gene_l=1.jpeg} }}%
  %\hfill
  %\subfloat[(iv) L=4]{{ \includegraphics[scale=0.08]{figures/bipartite_networks/drug_gene_l=3.jpeg} }}%
  %\caption{\textsl{Drug-Gene} ordered adjacency matrix for the learned $D=2$ embeddings of \textsc{HBDM-Re} for hierarchical tree structure of $L=4$. }\label{fig:drug_gene_dend}
%\end{figure}

%\begin{figure}
 % \centering
 % \subfloat[(i) \textsl{Drug-Gene}]{{ \includegraphics[scale=0.15]{figures/bipartite_networks/drug_genescatter_re.jpeg} }}%
 % \hfill
 % \subfloat[(ii) \textsl{GitHub}]{{ \includegraphics[scale=0.05]{figures/bipartite_networks/githubscatter_re.jpeg} }}%
  %\hfill
  %\subfloat[(iii) \textsl{Gottron-Reuters}]{{ \includegraphics[scale=0.05]{figures/bipartite_networks/reutersscatter_re.jpeg} }}%
  %\caption{ The $D=2$ learned latent space of \textsc{HBDM-Re} for various bipartite networks. }\label{fig:bip_sc}
%\end{figure}

%% file: 5-discussion.tex
\section{Discussion}
We developed the \textsc{HBDM}, a scalable reconciliation of latent distance models and their ability to account for homophily and transitivity with hierarchical representations of network structures.
%most of all we tried to define what constitutes a fine embedding approach for accurate graph representation. %We summarize the characteristics an embedding should express as: 1) Being interpretable by human perception, similar nodes are positioned closer in the latent space (one of the main goals and intuition behind GRL). 2) Providing hierarchical/multiresolutional intrinsic structures of the network facilitating interpretation and visualization at multiple scales. 3) Not depending on heuristic dimensionality reduction approaches but providing accurate low-dimensional representations with maximum $D=3$. 4) Showing good performance in the downstream tasks such as link prediction/network reconstruction/node classification.
We demonstrated how the proposed \textsc{HBDM} provides favorable network representations by: (1) Operating with a Euclidean distance metric providing an intuitive human perception of node similarity. (2) Naturally representing multiscale hierarchical structure based on its block structure and carefully designed clustering procedure optimized in terms of Euclidean distances. %to centroids within the latent space. 
(3) Directly and consistently operating in $D=2,3$ with high performance. (4) Performing well on all considered downstream tasks %(i.e., link-prediction, node classification, network visualization and community detection) 
highlighting its ability to account for the underlying network structure. Importantly, the inferred hierarchical structure admits community discovery at multiple scales as highlighted by the inferred dendrograms and ordered adjacency matrices, and naturally extends to the characterization of communities of bipartite networks.

%\textbf{Networks accurately embedded using ultra-low dimensions.}
Our finding of ultra-low dimensional accurate characterizations of network structures supports the findings in \cite{exact_Emb} in which a logistic PCA model was found to enable exact low-dimensional recovery of multiple real-world networks. % verifying that complex networks can be absolutely represented in low dimensional spaces.
Whereas the work of \cite{exact_Emb} focuses on exact network reconstruction we find that generalizable patterns can be well extracted in ultra-low dimensional representations with performance saturating after just $D=8$ dimensions for all networks considered. Whereas \cite{exact_Emb} found that their low-dimensional space did not perform well in classification tasks we observed strong node classification performance by the low-dimensional representations provided by HBDM. Importantly, for node classification, we observed better performance using KNN as opposed to simple linear classification based on logistic/multinomial regression typically used for node classification. This highlights that whereas most GRL works use linear classifiers there is no guarantee that the embedding space will be linearly separable and performance should therefore be compared to non-linear classifiers as they may provide more favorable performance as observed in this study.% As such, KNN naturally exploits Euclidean distances as defined by the latent space which aligning with the objective of the \textsc{HBDM} model.

%\textbf{HBDM beyond Euclidean distances.}
% Recently, the use of other distance metrics such as hyperbolic embeddings have been proposed and found to provide improved representation of multi-scale structure of networks. The presented HBDM naturally extends to other measures of distance and future work should explore how the HBDM can provide an explicit hierarchical and scalable framework also for more advanced LDM formulations such as ...
%Although many successful approaches to learning the continuous representations in Euclidean space have been proposed, 
Recent pioneering works \cite{mlg2017_6,NIPS2017_59dfa2df} have drawn significant attention of the research community by questioning the conventional embedding space preference. % as also reviewed for the LSM family \cite{LSM_geo}
It is well known that many real-world networks show power-law degree distribution, or they can consist of latent hierarchical inner structures. Therefore, Euclidean space might not always be appropriate to represent such complex network architectures. It might also require higher-dimensional spaces to show comparable performance in the GRL tasks. The works of \cite{mlg2017_6,NIPS2017_59dfa2df} demonstrated that hyperbolic spaces, such as the Poincare disk model, can provide substantial benefits over the Euclidean space. The presented \textsc{HBDM} model %learns node representations in a $D$-dimensional manifold, and it also introduces a framework allowing for 
naturally extends to other distance measures and %Although it relies on a multiscale block-structured design, 
future studies should explore how the \textsc{HBDM} can be extended to hierarchical representations beyond Euclidean geometry. 

Covariate information plays an important role in the outstanding performance of GRL methods and especially GNNs. In the current \textsc{LSM} literature, side information is accounted for by extra regressors in the logit/log link functions expressing the likelihood of a dyad being connected. %Consequently, additional information %($\bm{\mathcal{X}} \in\mathbb{R}^{N\times R} $)
%is not consistently projected in the latent space. 
Using the Mahalanobis distance imposing a block-diagonal covariance matrix (see supplementary), the proposed \textsc{HBDM} can naturally incorporate covariate information directly to the latent space and notably construct multi-scale structures via the enriched and concatenated embedding of the latent variables and the covariate information. Our analysis presently did not explore side information and this is also why we did not include comparisons to prominent GNN-based approaches as these procedures do not provide favorable performance when only learning from the graph structure itself. As such, we observed (not shown) poor performance of GraphSage \cite{hamilton2017inductive} when only having access to the graph structure in the present setup.
%In more detail, we can define a new embedding matrix $\bm{\Bar{Z}}$ as the concatenation over the latent variables and the covariate information for node $i$ as: $\bm{\Bar{z}}_i=[\bm{z}_i;\bm{x}_i]$ and a Mahalanobis correlation matrix as: $\bm{S}=\begin{bmatrix} \bm{I} & \bm{0}\\ \bm{0}^T & \bm{J} \end{bmatrix} \in\mathbb{R}^{(D+R)\times (D+R)}$, where $I \in\mathbb{R}^{D\times D}$ the identity matrix, the zero matrix $\bm{0} \in\mathbb{R}^{D\times R}$ and the covariate coefficient matrix $\bm{J} \in\mathbb{R}^{R\times R}$. In this setting, \textsc{HBDM} is able to construct a covariate information-aware multi-scale latent space by the use of the Mahalanobis distance $d_{ij}=\sqrt{(\bm{\Bar{z}}_i-\bm{\Bar{z}}_j)^T\bm{S}^{-1}(\bm{\Bar{z}}_i-\bm{\Bar{z}}_j)}$.
The \textsc{HBDM} operates on static networks and thus is not naturally an inductive model. Nevertheless, potential new emerging nodes can be projected into the inferred latent space by fixing the embeddings of  nodes present in the training set while optimizing the new nodes for their locations in the learned latent space. We leave a comparison of such a strategy against naturally inductive models such as GNNs for future work. 

%\textbf{Hierarchical multi-scale structures.} 
Our discoveries highlight the existence and importance of hierarchical multi-scale structures in complex networks. The across hierarchy re-ordered adjacency matrices given by \textsc{HBDM}, manifest sub-communities inside of what already appears as a strongly connected community. This points to how delicate the task of defining communities is and the importance of accounting for communities at multiple scales, as enabled by the \textsc{HBDM}. Importantly, these results generalize for bipartite networks where multi-scale geometric representations, joint hierarchical structures, and community discovery are arduous tasks.

%\textbf{Limitations in regards to stochastic equivalence.}
The \textsc{HBDM} uses the \textsc{LDM} and thus is good at characterizing transitivity and homophily at a node and cluster level, whereas the random effects enable accounting for degree heterogeneity. Notably, the \textsc{HBDM} suffers from the limitations of the \textsc{LDM} and is thus unable to model stochastic equivalence. Future work should therefore investigate hierarchical structures imposed on more flexible GRL procedures enabling stochastic equivalence and contrast the performance when accounting for stochastic equivalence to the existing hierarchical methods based on the \textsc{SBM} \cite{clauset2008hierarchical, roy2007learning, herlau2012detecting,agglo_bayes, herlau2013modeling,Peixoto_2014}.

In conclusion, we proposed the Hierarchical Block Distance Model (\textsc{HBDM}), a scalable reconciliation of network embeddings using the latent distance model (\textsc{LDM}) and hierarchical characterizations of structure at multiple scales via a novel clustering framework. Notably, the model mimics the behavior of the \textsc{LDM} where the use of homophily and transitivity is most important while scaling in complexity by $\mathcal{O}(DN\log{N})$. We analyzed thirteen networks from moderate sizes to large-scale with the \textsc{HBDM} having favorable performance when compared to existing scalable embedding procedures. In particular, we observed that the \textsc{HBDM} well predicts links and node classes %utilizing a very low embedding dimension of $D=2$ 
providing accurate network visualizations and characterization of structure at multiple scales. Our results demonstrate that favorable performance can be achieved using ultra-low (i.e. $D=2$) embedding dimensions and a scalable hierarchical representation that accounts for homophily and transitivity.